\documentclass[aps,prd,reprint,superscriptaddress,longbibliography,showpacs]{revtex4-2}

\usepackage{graphicx}% Include figure files
\usepackage{dcolumn}% Align table columns on decimal point
\usepackage{bm}% bold math
\usepackage[figurename=Figure]{caption}
\usepackage{float}    % For tables and other floats
\usepackage{verbatim} % For comments and other
\usepackage{amsmath}  % For math
\usepackage{upgreek} 
\usepackage{amssymb}  % For more math
\usepackage{listings} % For source code
\usepackage{graphics}
\usepackage{enumitem} % useful for itemization
\usepackage{latexsym,epsfig}
\usepackage{subcaption}
\usepackage{stackrel}

\usepackage{placeins} %This contains \FloatBarrier command.

\usepackage[colorlinks=true,breaklinks=true,allcolors=blue]{hyperref}
\usepackage{lipsum}
\usepackage{dsfont}
\usepackage[scr=boondox]{mathalfa}

\def\Oo{\ensuremath{{\cal O}}} %Order sign
\def\Jj{\ensuremath{{\cal J}}} %effective coupling
\def\Ll{\ensuremath{{\cal L}}}

\def\Ss{\ensuremath{{\cal S}}} 
\def\Hh{\ensuremath{{\cal H}}} %the Hamiltonian
\def\Nn{\ensuremath{{N}}}

\def\Gg{\ensuremath{{\cal G}}} % Green's functions
\def\Zz{\ensuremath{{\cal Z}}}

\def\Qq{\ensuremath{{\cal Q}}}

\usepackage[scr=boondox]{mathalfa}

\def\hh{\ensuremath{\mathscr{h}}}

\def\p{\ensuremath{{\partial}}}
\def\i{\ensuremath{{\imath}}}

\def\Im{\ensuremath{{\operatorname{Im}}}}
\def\Tr{\ensuremath{{\operatorname{Tr}}}}
\def\Re{\ensuremath{{\operatorname{Re}}}}
\def\p{\ensuremath{{\partial}}}
\def\betat{\ensuremath{{\tilde{\beta}}}}
\def\Qqt{\ensuremath{{\tilde{\mathcal{Q}}}}}
\def\ttil{\ensuremath{{\tilde{T}}}}
\def\mut{\ensuremath{{\tilde{\mu}}}}
\def\Omegat{\ensuremath{{\tilde{\Omega}}}}
\def\tc{\ensuremath{{\tilde{T}_c}}}
\def\qc{\ensuremath{{\tilde{\mathcal{Q}}_c}}}
\def\muc{\ensuremath{{\tilde{\mu}_c}}}

\def\qtil{\ensuremath{{\tilde{\mathcal{Q}}}}}
\newcommand\ex[1]{ \langle #1 \rangle }

\newcommand{\be}{\begin{equation}}
	\newcommand{\ee}{\end{equation}}
\newcommand{\bea}{\begin{equation}\begin{aligned}}
		\newcommand{\eea}{\end{aligned}\end{equation}}
\newcommand{\ben}{\begin{enumerate}}
	\newcommand{\een}{\end{enumerate}}

\DeclareDocumentCommand{\nint}{ O{} O{} m }{\ensuremath{ \int_{\mbox{\scriptsize $#1$}}^{\mbox{\scriptsize$#2$}}\!\!\! \mbox{\small $\,\mathrm{d}#3$\! }}}

\usepackage{tikz,bm} % Useful for drawing plots
\usepackage{circuitikz}
\usetikzlibrary{decorations.markings}
\usetikzlibrary{decorations.pathreplacing,calligraphy}

\definecolor{mycolor}{rgb}{1,0.2,0.3}
\definecolor{brightgreen}{rgb}{0.4, 1.0, 0.0}
\definecolor{britishracinggreen}{rgb}{0.0, 0.26, 0.15}
\definecolor{cadmiumgreen}{rgb}{0.0, 0.42, 0.24}
\definecolor{ceruleanblue}{rgb}{0.16, 0.32, 0.75}
\definecolor{darkelectricblue}{rgb}{0.33, 0.41, 0.47}
\definecolor{darkpowderblue}{rgb}{0.0, 0.2, 0.6}
\definecolor{darktangerine}{rgb}{1.0, 0.66, 0.07}
\definecolor{emerald}{rgb}{0.31, 0.78, 0.47}
\definecolor{palatinatepurple}{rgb}{0.41, 0.16, 0.38}
\definecolor{pastelviolet}{rgb}{0.8, 0.6, 0.79}

\begin{document}
	
	\preprint{APS/123-QED}
	
	\title{Thermodynamics and dynamics of coupled complex SYK models}% Force line breaks with \\
	%\thanks{A footnote to the article title}%
	
	\author{Jan C. Louw}
	%\altaffiliation[Also at ]{Physics Department, XYZ University.}%Lines break automatically or can be forced with \\
	\email{jan.louw@theorie.physik.uni-goettingen.de}
	\affiliation{%
		Institute for Theoretical Physics, Georg-August-Universit\"{a}t G\"{o}ttingen, Friedrich-Hund-Platz 1, 37077 G\"ottingen, Germany
	}%
	\author{Linda M. van Manen}
	\email{linda.van.manen@uni-jena.de}
	\affiliation{%
		Friedrich-Schiller-Universit\"at, Institute for Theoretical Physics, Max‐Wien‐Platz 1, 07743 Jena, Germany
	}%
	\author{Rishabh Jha}
	\email{rishabh.jha@uni-goettingen.de}
	\affiliation{%
		Institute for Theoretical Physics, Georg-August-Universit\"{a}t G\"{o}ttingen, Friedrich-Hund-Platz 1, 37077 G\"ottingen, Germany
	}%

	%\collaboration{MUSO Collaboration}%\noaffiliation
	
	%\author{Charlie Author}
	% \homepage{http://www.Second.institution.edu/~Charlie.Author}
	%\affiliation{
		% Second institution and/or address\\
		% This line break forced% with \\
		%}%
	%\affiliation{
		% Third institution, the second for Charlie Author
		%}%
	%\author{Delta Author}
	%\affiliation{%
		% Authors' institution and/or address\\
		% This line break forced with \textbackslash\textbackslash
		%}%
	
	%\collaboration{CLEO Collaboration}%\noaffiliation
	
	%\date{\today}% It is always \today, today,
	%  but any date may be explicitly specified
	
	% Here we extend the work and consider a system of coupled large-$q$ SYK model and shows that despite having an additional scale in the system thereby leading to different equations for equation of state and grand potential compared to the single case, we still obtain the same universality class for our coupled model as that of (mean-field) van der Waals. We further show that the system remains maximally chaotic to leading order in $1/q$ despite the presence of an additional scale in the system. Therefore the thermodynamic properties around the critical point as well as the dynamical Lyapunov exponents are identically the same as that of the single model. 
	
	\begin{abstract}
		
		It has been known that the large-$q$ complex SYK model falls under the same universality class as that of van der Waals (mean-field) and saturates the Maldacena-Shenker-Stanford bound, both features shared by various black holes. This makes the SYK model a useful tool in probing the fundamental nature of quantum chaos and holographic duality. This work establishes the robustness of this shared universality class and chaotic properties for SYK-like models by extending to a system of coupled large-$q$ complex SYK models of different orders. We provide a detailed derivation of thermodynamic properties, specifically the critical exponents for an observed phase transition, as well as dynamical properties, in particular the Lyapunov exponent, via the out-of-time correlator calculations. Our analysis reveals that, despite the introduction of an additional scaling parameter through interaction strength ratios, the system undergoes a continuous phase transition at low temperatures, similar to that of the single SYK model. The critical exponents align with the Landau-Ginzburg (mean-field) universality class, shared with van der Waals gases and various AdS black holes. Furthermore, we demonstrate that the coupled SYK system remains maximally chaotic in the large-$q$ limit at low temperatures, adhering to the Maldacena-Shenker-Stanford bound, a feature consistent with the single SYK model. These findings establish robustness and open avenues for broader inquiries into the universality and chaos in complex quantum systems. We provide a detailed outlook for future work by considering the ``very" low-temperature regime, where we discuss relations with the Hawking-Page phase transition observed in the holographic dual black holes. We present preliminary calculations and discuss the possible follow-ups that might be taken to make the connection robust.
		
	\end{abstract}
	
	%\keywords{Suggested keywords}%Use showkeys class option if keyword
	%display desired
	\maketitle
	
		\noindent \textbf{Corresponding Author:} Rishabh Jha 
	%\tableofcontents

	\section{Introduction}
	\label{introduction}
	
	The Sachdev-Ye-Kitaev (SYK) model, initially conceptualized in the context of condensed matter physics, has rapidly emerged as a pivotal tool in exploring quantum gravity and chaos. Characterized by its non-Fermi liquid behavior and lack of quasiparticle excitations, the SYK model exemplifies a quantum many-body system that is exactly solvable in certain limits. The large-$q$ complex SYK model, in particular, has been instrumental in revealing connections between quantum chaos, black hole thermodynamics, and the holographic principle. We define quantum chaos in terms of exponential growth in the out-of-time-correlators (OTOCs). Studies have proved that the SYK model exhibits maximal chaos, saturating the Maldacena-Shenker-Stanford (MSS) bound, and parallels the thermodynamic properties of black holes, particularly in the context of the holographic duality \cite{Maldacena2016Nov,Kitaev2015,Louw2023Oct}.
	
	It was recently shown that the large-$q$ SYK model has a second order phase transition, which matches the universality class of a wide range of black holes \cite{Louw2023Feb}. Similar phase transitions are also found in the finite-$q$ case \cite{Azeyanagi2018Feb,Ferrari2019Jul}. Given the analytic solvability of the large-$q$ model, one was able to find the exact thermodynamics and find its exact holographic dual gravitation model \cite{Louw2023Oct}. This duality is in terms of the thermodynamics. However, there are also large overlaps in the dynamics, namely the Lyapunov exponents, of the respective models. Here the thermodynamics is in the grand canonical ensemble. On the SYK side, this means that we have a chemical potential coupled to the charge density. The phase transition results from a varying conjugate field (chemical potential) inducing a jump in charge density. The analogy to gravity is most easily seen in a standard $3+1$-dimensional asymptotically Anti-de-Sitter (AdS$_{3+1}$) black hole system, where the black hole charge is the conjugate field and the surface charge density is the order parameter (see analogy 2 in \cite{Kubiznak2012Jul}). Indeed, these two parameters are mapped onto one another in the thermodynamic dictionary between the $0+1$-dimensional SYK dot and the AdS$_{1+1}$ charged black hole system \cite{Louw2023Oct}. The jump in charge densities of the SYK dot is then directly related to a jump in black hole density. In particular, since the SYK dot has no size, the charge density (charge per flavors of fermions) jump actually corresponds to a change in charge. On the black hole side, the charge is merely a conjugate field, i.e., it is a set parameter and the jump happens in the surface charge density due to the black hole transition between small and large horizon radii.
	
	For all cases, the coexistence line between the two phases terminates at the critical point, e.g., $(\mu_c,T_c)$ for the SYK model \cite{Louw2023Feb} or $(Q_{c},1/T_c)$ for AdS$_{3+1}$ (Fig. 13 in \cite{Kubiznak2012Jul}). These critical points correspond to the van der Waals second order phase transition (Landau-Ginzburg mean field critical exponents) \footnote{Landau-Ginzburg universality class implies having the same critical exponents corresponding to a second-order phase transition as predicted by the Landau-Ginzburg theory, namely $\alpha = 0$, $\beta = 1/2$, $\gamma = 1$ and $\delta = 3$. We have defined these exponents later in this work when we derive them for our model. van der Waals fluids, AdS black holes and other mean-field theories satisfy the same and therefore, ``Landau-Ginzburg universality class'', or ``van der Waals universality class'', or ``mean-field universality class'' are all the same and interchangeably used in the literature.}. Note the difference that the SYK model has no transition at temperatures above $T_c$, while the AdS$_{3+1}$ case has no transition for temperatures below $T_c$. Both transitions also have a non-trivial end-point corresponding to a first order chaotic-to-regular phase transition. 
	
	There is a caveat, though. Only for the uncharged case does the smaller black hole evaporate, such that we are left with regular (non-chaotic) non-interacting thermal radiation for a particular value of temperature \cite{Hawking1982Jan}. On the SYK side, however, there exists a whole range of temperatures where we have a chaotic to regular (non-chaotic phase transition) \cite{Louw2023Feb} (see section \ref{very low temperatures} below). Although the thermodynamics between the SYK model and thermodynamically holographic dual does match, there is a short-coming of this duality in that it does not reproduce the standard Hawking-Page transition \cite{Hawking1982Jan}. We make this point precise as outlook for future work in sections \ref{very low temperatures} and \ref{hawking page section} below. We provide preliminary calculations in the outlook and highlight what needs to be done as a future work to make this connection robust, if at all possible. But at the same time, the SYK model has proved instrumental in exploring connections between quantum chaos and black holes \cite{Cotler2017May, Chowdhury2022Sep}. The natural question to ask is how robust are these quantum chaotic and thermodynamic properties of \textit{SYK-like} models.
	
	In this paper, we study the robustness of the chaotic and thermodynamic properties of a system of coupled large-$q$ complex SYK models and the associated dynamical inequivalence to black holes in the context of Hawking-Page transition. We do this by considering the sum of two non-commuting SYK models of different orders. This has the effect of introducing an additional scale. As expected from relevance, in a renormalization context, the lower order term dominates at low temperatures. Since an action consists of a kinetic and a potential term, a strongly dominating kinetic term results in a free, hence non-chaotic, system. As such, it seems reasonable, that such a term would have a large effect on the thermodynamics. Here, interactions are of order $q$, while we add a pseudo-kinetic term of order $q/2$. The term ``pseudo-kinetic'' is motivated by various findings. For one, such a term coupling a lattice of dots at equilibrium yields the standard intermediate temperature resistivity related to the $q=4$, hence quadratic hopping $q/2=2$, case \cite{Patel2018Oct,Cha2020Sep}. This is in agreement with the observation that the large-$q$ expansions converge rather fast, leading to a large quantitative and qualitative overlap even with the $q=4$ case as discussed at length in \cite{Louw2023}. This then gives us the advantage of an analytically solvable system via a $1/q$ expansion, as opposed to the $q=4$ case which is only numerically solvable. This is our motivation in calling the $q/2$ term in our model below as ``pseudo-kinetic'' term because of this convergence and qualitative similarity with the finite $q$ case, such as $q=4$ while retaining the analytical tractability of the system. However, we note that since the $q/2$ model is also chaotic for values larger than four, as opposed to the non-chaotic kinetic ($q=2$) SYK model; therefore we expect the term ``pseudo-kinetic'' to be less appropriate at low temperatures. Another difference that occurs is that the large-$q$ case thermalizes instantly \cite{Louw2022Feb} while the $q=4$ case thermalizes at the Planckian rate \cite{Eberlein2017Nov}. The Majorana variant (corresponding to the uncharged case) of our model was considered in \cite{Jiang2019Aug}. However, as we will see in this work that including a $U(1)$ symmetry yields a much richer phase diagram where we observe a phase transition at low-temperatures.
	
	\subsection{Results}
	Since the two Hamiltonians do not commute, one might expect rather different thermodynamics. The addition of a new scaling parameter, the ratio of interaction strengths, adds a layer of complexity to the model.  Despite this, we find that the system retains the characteristic continuous phase transition of single large-$q$ SYK model for all ratios in coupling strength and the phase transition persists in the low-temperature regime. The critical point shifts for each value of the ratio of the respective coupling strengths. However, we find that this variant of coupled large-$q$ SYK models in fact has the same partition function around the critical point. In other words, the coupling of two SYK models still flows to the same model under renormalization. The critical exponents observed in this coupled system intriguingly align with the Landau-Ginzburg (mean-field) universality class, which is a shared feature with van der Waals gases and a spectrum of AdS black holes \cite{Kubiznak2012Jul}. This finding underscores a potentially universal behavior underlying these disparate physical systems. 
	
	Regarding the chaotic properties in our system of coupled SYK models, it seems reasonable that the introduction of a new scale in the system, namely the ratio of coupling constants of the two SYK terms in the Hamiltonian, might change the dynamics at lower temperatures. This would be reflected in some novel coupling found between the $q$ and $q/2$ related kernels defining the time evolutions of the OTOCs. Surprisingly, we find that, like the self energies \cite{Maldacena2016Nov}, the kernels are also additive for our coupled SYK system. The chaotic nature does not change much, since the dominating SYK term just leads to the same maximal chaos as the single SYK term in the low-temperature limit. As such, we find that the dynamics of our coupled SYK system has a rather similar phase diagram to that of the single SYK model. This retention of maximal chaos, as denoted by adherence to the MSS bound, aligns with the behavior observed in single large-$q$ complex SYK model \cite{Maldacena2016Nov}. The persistence of maximal chaos in the coupled system is not just a continuation of the SYK model's established traits but also highlights the robust nature of chaos in these quantum systems. This work establishes this robustness for a coupled SYK system not just for the case of chaos but also for the shared universality with the Landau-Ginzburg mean field critical exponents.

	We have provided a detailed section on conclusion and outlook (section \ref{conclusion section}) where we have shown preliminary calculations to continue our analysis at further ``\emph{very}'' low-temperatures where we observe a first order phase transition between chaotic and non-chaotic phases (see Fig. \ref{phase diagram for all T} for the phase diagram). We compare this with the standard Hawking-Page phase transition observed in the holographic dual black holes. We clearly highlight the assumption involved in these calculations in the outlook and leave them as future work.

	We introduce the model and the associated setup in Section \ref{model section} where we present the exact solutions of Green's functions. Section \ref{eos section} discusses the thermodynamics of our model, where we evaluate the equation of state as well as the grand potential. We observe a continuous phase transition at low-temperature. We analyze the associated critical point in section \ref{universality} where we explicitly evaluate the critical exponents and find them to belong to the same universality class as Landau-Ginzburg mean-field exponents, a universality class also shared by van der Waals fluids and various AdS black holes. Then we analyze the dynamics of our model in section \ref{LyapunSec} where we study the out-of-time correlator (OTOC) and calculate the Lyapunov exponent. We find that our system is chaotic and saturates the MSS bound at low temperatures. We also discuss the relation to the phase transition at ``very'' low-temperature (we define therein what ``very'' low temperatures are) and the associated comparison with the Hawking-Page phase transition. We conclude and provide an outlook in section \ref{conclusion section}.

	\section{Model}
	\label{model section}
	
	We consider the following coupled complex large-$q$ SYK system Hamiltonian in equilibrium:
	\begin{equation}
		\Hh =  J_q\Hh_q + J_{q/2} \Hh_{q/2}
		\label{model}
	\end{equation}
	where $\Hh_{\kappa q}$ ($\kappa \in \{ \frac{1}{2}, 1\}$) is large-$q$ complex SYK model Hamiltonian given by
	\begin{equation}
		\Hh_{\kappa q}= \hspace{-1mm} \sum\limits_{\substack{ \{\bm{\mu}\}_1^{\kappa q/2} \\ \{\bm{\nu}\}_1^{\kappa q/2} }}^N \hspace{-2mm} X^{\bm{\mu}}_{\bm{\nu}} c^{\dag}_{\mu_1} \cdots c^{\dag}_{\mu_{\kappa q/2}} c_{\nu_{\kappa q/2}}^{\vphantom{\dag}} \cdots c_{\nu_1}^{\vphantom{\dag}} 
	\end{equation}
	summing over $\{\bm{\mu}\}_{1}^{\kappa q/2} \equiv 1\le \mu_1<\cdots< \mu_{\kappa q/2}\le\Nn$. Here $X^{\bm{\mu}}_{\bm{\nu}}$ is a random matrix whose components are derived from a Gaussian ensemble with zero mean and variance given by
	\begin{equation}
		\overline{|X|^2} =  \frac{(\kappa q)^{-2}((\kappa q / 2) !)^2}{(N / 2)^{\kappa q-1}} 
		\label{variance of X random matrix}
	\end{equation}
	
	Our main focus will be the thermodynamics of this model, defined by the partition function $\Zz = e^{-\beta N \Omega}$.  The thermodynamics of this model is, in fact, equivalent to a uniformly coupled lattice of SYK dots with $q/2$-body hopping with $q$-body inter-dot interactions as considered in \cite{Jha2023}. We have proved this equivalence in appendix \ref{equivalence of syk chain with syk dot}, justifying calling our model in Eq. \eqref{model} as coupled SYK model. Here the grand potential (per lattice site) $\Omega$  may, in fact, be found by studying the Green's functions which are defined as follows:
	\begin{equation}
		\mathcal{G}\left(t_1, t_2\right) \equiv \frac{-1}{N} \sum_{\mu=1}^N\left\langle\mathcal{T}_{\mathcal{C}} c_{ \mu}\left(t_1\right) c_{ \mu}^{\dagger}\left(t_2\right)\right\rangle
	\end{equation}
	where $\mathcal{T}_{\mathcal{C}}$ is the time ordering operator on the Keldysh contour \cite{Kamenev2009May}. This Green's function may be separated into the greater ($t_1>t_2$) and lesser ($t_2>t_1$) Green's functions, which depict the order of the operators. We use the ansatz for the Green's functions of large-$q$ complex SYK model \cite{Louw2022, Louw2023Feb, Jha2023}
	\begin{equation}
		\Gg^{\gtrless}(t_1,t_2) = (\Qq \mp 1/2) e^{g^{\gtrless}(t_1,t_2)/q} \label{GreenDef1}.
	\end{equation}
	where the system has a conserved total charge per flavor defined by
	\begin{equation}
		\mathcal{Q} \equiv \frac{1}{N} \sum_{\alpha=1}^N\left[c_{\alpha}^{\dagger} c_{\alpha}-1 / 2\right].
		\label{local charge density definition}
	\end{equation}
	and $g^{\gtrless}$ satisfies the boundary conditions
	\begin{equation}
		g^{\gtrless}(t,t) = 0.
	\end{equation}
	
	Before proceeding further, we introduce some notations introduced earlier in the literature \cite{Louw2022, Louw2023Feb, Jha2023} that will serve us later in our analysis. Our main focus of interest will be in the so-called ``symmetric'' and ``asymmetric'' little $g$ defined as follows:
	\begin{equation}
		g^{\pm}(t_1,t_2) \equiv \frac{g^>(t_1,t_2) \pm g^{<}(t_2,t_1)}{2}.
		\label{def of g plus minus}
	\end{equation}
	Notice the reverse time argument in $g^<$ on the right-hand side where we recall that $g^{\gtrless}\left(t_1, t_2\right)^*=g^{\gtrless}\left(t_2, t_1\right)$. We introduce the effective coupling coefficients for both the terms in the Hamiltonian in Eq. \eqref{model}
	\begin{equation}
		\Jj_{\kappa q} \equiv (1-4\Qq^2)^{\kappa q/4 - 1/2}J_{\kappa q}
		\label{effective coupling0}
	\end{equation}
	where again $\kappa \in \{ \frac{1}{2}, 1\}$. 
	
	We defer the derivation of differential equations for $g^{\pm}$ to appendix \ref{derivation of differential equations for gs}. Recall that we are only interested in the equilibrium situation, implying that the Green's functions are only dependent on time differences, $t_{12} \equiv t_1-t_2$, i.e., $g^{\pm}(t_1,t_2) \equiv g^{\pm}(t_{12})$. Using appendix \ref{derivation of differential equations for gs}, we get as solution for the Green's function $g^-$ as follows:
	\begin{equation}
		g^{-}(t_1,t_2)= g^{-}(t_{1}-t_{2})= g^{-}(t_{12})=2\Qq \i \alpha \,t_{12}. 
	\end{equation}
	Here, using Eq. \eqref{first order DE for g}, we see that $\alpha$ is related to the derivative of the symmetric part
	\begin{equation}
		\dot{g}^+(0) = \i \alpha \label{alphaG+}
	\end{equation}
	This symmetric part $g^+(t_{12})$ is the solution to a second order differential equation. Since we only have to focus on time differences due to equilibrium condition, we focus on the Green's function with single time argument which we label $t$
	\begin{equation}
		\ddot{g}^{+}(t) = -2\Jj_q^2 e^{g_+(t)} - 2\Jj_{q/2}^2 e^{g_+(t)/2}
	\end{equation}
	Fortunately, this differential equation can be solved in closed form and the solution is given by
	\begin{equation}
		e^{g^{+}(t) / 2}=\frac{1}{\left(\beta \mathcal{J}_{q / 2}\right)^2} \frac{(\pi v)^2}{1+\sqrt{A^2+1} \cos (\pi v(1 / 2-\imath t / \beta))}
		\label{solution for g}
	\end{equation}
	\cite{Eberlein2017Nov}, with the dimensionless parameter
	\begin{equation}
		A \equiv \frac{\pi v \beta \mathcal{J}_q}{\left(\beta \mathcal{J}_{q / 2}\right)^2}.
		\label{A}
	\end{equation}
	
	We impose the boundary condition for $g$, namely $g^+(0) = 0$ to get the closure relation for $v$ as follows:
	\begin{equation}
		\pi v=\sqrt{\left(\beta \mathcal{J}_q\right)^2+\left(\frac{\left(\beta \mathcal{J}_{q / 2}\right)^2}{\pi v}\right)^2} \cos (\pi v / 2)+\frac{\left(\beta \mathcal{J}_{q / 2}\right)^2}{\pi v}
		\label{closure}
	\end{equation}
	
	Using the first equation in Eq. \eqref{first order DE for g}, we see that at equal times, we have
	\begin{equation}
		\dot{g}^+(0) = \i \alpha
	\end{equation}
	where $\alpha$ is a constant in equilibrium whose details can be found in appendix \ref{derivation of differential equations for gs}. For convenience, we reproduce the definition of $\alpha$ here from Eq. \eqref{alpha energy}. $\alpha$ is related to the expectation value of energy per particle given by
	\begin{equation}
		(1-\Qq^2)\alpha(t) = \epsilon_{q/2}(t)/2 + \epsilon_{q}(t)
	\end{equation}
	where in the large-$q$ limit, we have $\Qq \sim \Oo(\frac{1}{\sqrt{q}})$ and $\epsilon_{\kappa q}(t) \equiv q^2\frac{J_{\kappa q}\ex{\Hh_{\kappa q}}}{N} $ where $\kappa \in \{\frac{1}{2}, 1 \}$.
	
	This allows us to explicitly calculate $\alpha$ using the solution for $g^+$ in Eq. \eqref{solution for g} to get
	\begin{equation}
		\beta\alpha = -2 \sqrt{\left( \frac{\left( \beta\Jj_{q/2} \right)^4}{\pi^2 v^2} + \left( \beta\Jj_q \right)^2 \right)} \sin(\pi v/2)
		\label{alpha}
	\end{equation}
	where we have made use of the closure relation Eq. \eqref{closure}.

	\section{Equation of State and Grand Potential}
	\label{eos section}
	
	We start by calculating the equation of state valid for all temperatures, and then identifying the region of interest where we can observe a phase transition. Following the logic in \cite{Louw2022, Louw2023Feb}, we impose the Kubo-Martin-Schwinger (KMS) relation \cite{Stefanucci2013Mar} on the full Green's function of the system that leads us to the equation of state for our system given by
	\begin{equation}
		\beta\mu =  \ln\left[\frac{1+2\Qq}{1-2\Qq}\right] - 2 \Qq \beta\alpha/q \label{EOS}
	\end{equation}
	We immediately notice that if $\beta \alpha = \Oo(q^0)$ and $\Qq = \Oo(q^{-1/2})$, the second term will be suppressed in large-$q$ limit. Therefore, at intermediate and large temperatures, we are left with the equation of state of a free system only. But if we rescale the temperature as $\betat = \beta / q$ such that $\betat \sim \Oo(q^0)$, then we allow for the second term in the equation of state to contribute as well. Given this re-scaling, we are accordingly interested in the low-temperature regime as we are considering the large-$q$ limit. 
	
	Note that the effective interactions are exponentially suppressed as $\Jj_{\kappa q} \sim e^{-\kappa q \Qq^2}$ (using Eq. \eqref{effective coupling0}) for non-rescaled charge densities. As such, our first step considers charge densities 
	\begin{equation}
		\Qq = \frac{\tilde{\Qq}}{q^{1/2}}, \qquad \tilde{\Qq} = \Oo(q^0)
		\label{scaling1}
	\end{equation}
	As noted above, we also focus on the low-temperature regime 
	\begin{equation}
		T = \ttil/q = \Oo(q^{-1}),\quad \beta = q \tilde{\beta} \equiv q/\ttil
		\label{scaling2}
	\end{equation}
	and a rescaled chemical potential 
	\begin{equation}
		\mu = \tilde{\mu} q^{-3/2}
		\label{scaling3}
	\end{equation}
	where there is a competition between the interaction terms and the free system contribution. To better understand the equation of state in the low-temperature regime, we need to first find $v$ using Eq. \eqref{closure} in that limit.
	
	\subsection{Finding v in the low-temperature limit}
	We must find the solution $v$ to Eq. \eqref{closure} in the low-temperature regime. In the case where we set $J_{q/2}=0$, we would be left purely with $v\sim 1$. This changes here, however. Let us define 
	
	\begin{equation}
		U \equiv \tilde{\beta} \Jj_{q}, \qquad K  \equiv \tilde{\beta} \Jj_{q/2}
		\label{U and K defs}
	\end{equation}
	leaving 
	\begin{equation}
		A = \frac{\pi v U}{q K^2} = \Oo(q^{-1}). \label{A2}
	\end{equation}
	With this, the Eq. \eqref{closure} takes the form
	\begin{equation}
		\pi v = \sqrt{ \left( \frac{U^2}{q^2} + \left(\frac{K^2}{\pi v}\right)^2 \right)} q^2 \cos(\pi v/2) + \frac{q^2 K^2}{\pi v}, 
		\label{closure2}
	\end{equation}
	where we can interpret $U$ as an interaction term while $K$ as a pseudo-kinetic term. We take both of the dimensionless couplings to be of order $q^0$. From the above we note how the pseudo-kinetic term, corresponding to $q/2$-body interactions, is the relevant term, since it enters in at a higher order of $q$. If we however first set $K=0$, then this would yield the standard closure relation \cite{Maldacena2016Nov} $\pi v = U q \cos(\pi v/2)$ solved by $v = 1 +\Oo(1/q)$. Since this special case has already been treated extensively \cite{Louw2023Feb,Louw2023Oct}, we focus on the case of non-zero $K$ at large-$q$
	\begin{equation}
		\pi v = \left[\frac{K^2}{\pi v} + \frac{U^2 \pi v}{2q^2 K^2} +\Oo(1/q^4) \right] q^2 \cos(\pi v/2) + \frac{q^2 K^2}{\pi v}, 
	\end{equation}
	with the solution
	\begin{equation}
		v=2 - \frac{4}{q K} \sqrt{2 + (U/K)^2} + \Oo(q^{-2}) .
		\label{vSol}
	\end{equation}
	We realize that for our model, $v \in [0,2]$. We know that in the $K=0$ case that the low-temperature limit would yield $v\to 1$ \cite{Maldacena2016Nov}. We will return to this doubling of $v$ in \eqref{vSol} from the $K=0$ case when we discuss Lyapunov exponents in Section \ref{LyapunSec}. 
	
	\subsection{Equation of state}
	
	Since we have $\alpha$ in the equation of state (Eq. \eqref{EOS}), we make use of Eq. \eqref{vSol} together with Eq. \eqref{alpha} to get $\alpha$ in the low-temperature limit as follows: 
	\begin{equation}
		\beta\alpha/q = -2 \sqrt{\frac{q^2 K^4}{\pi^2 v^2} +  U^2} \sin(\pi v/2) \sim - 2 \sqrt{2 K^2 + U^2}.
		\label{alpha at low T}
	\end{equation}

	Then using the scaling of different quantities mentioned in eqs. \eqref{scaling1}, \eqref{scaling2} and \eqref{scaling3}, we are left with the equation of state in the low-temperature limit
	\begin{equation}
		\tilde{\beta}\tilde{\mu} =  4\tilde{\Qq}\Big[1+ \sqrt{2 K(\Qq)^2 + U(\Qq)^2}\Big] 
		\label{EOS-lowT}
	\end{equation}
	which notably reduces to the standard large-$q$ complex SYK model for $K=0$ \cite{Louw2023Feb} as it should.
	
	We now consider the equation of state where we first substitute from the respective definitions (namely, eqs. \eqref{effective coupling0} and \eqref{U and K defs} in large-$q$ scaled temperature limit) 
	\begin{equation}
		U = \betat J_q \exp \left(-\Qqt^2\right), K =\betat J_{q/2}  \exp \left(-\frac{\Qqt^2}{2}\right),
		\label{U and K defs2}
	\end{equation}
	to get
	\begin{equation}
		\frac{\mut}{J_q} = 4  \tilde{\Qq} \left[ \sqrt{2 \left(\frac{J_{q/2}}{J_q} \right)^2 e^{-\Qqt^2 }+  e^{-2\Qqt^2}} + \ttil/J_q\right]\label{muFull}
	\end{equation}
	
	We introduce a new dimensionless parameter of the system due to the new scale coming from the other SYK term compared to single large-$q$ complex SYK model, namely
	\begin{equation}
		\gamma \equiv \frac{J_{q/2}}{J_q} 
		\label{def of gamma}
	\end{equation}
	which makes our equation of state as
	\begin{equation}
		\frac{\mut}{J_q} = 4  \tilde{\Qq} \left[ \sqrt{2 \gamma^2 e^{-\Qqt^2 }+  e^{-2\Qqt^2}} + \ttil/J_q\right]\label{muFull}
	\end{equation}

\begin{figure}
		\centering
		\includegraphics[width=\columnwidth]{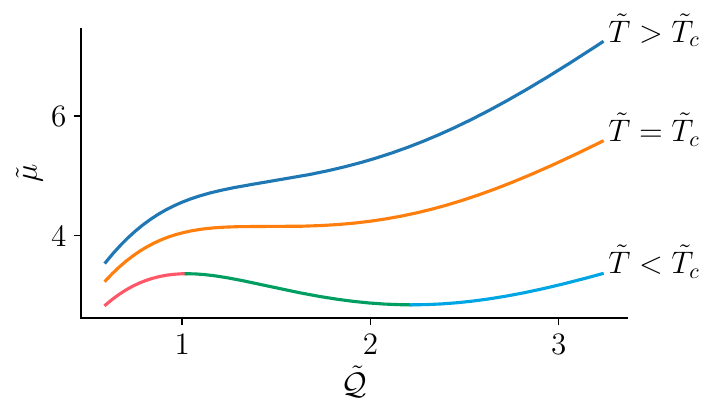}
		\caption{Plots of the equation of state (Eq. \eqref{muFull}) for various temperatures for $\gamma^2 = 0.28$ and $J_q = 1$. The blue and orange curves are both one-to-one functions, each corresponding to the supercritical phase at $T\ge T_c$. The three colors in the bottom curve correspond to $3$ different phases. Pink for the low density gaseous phase, green for the unstable phase, and blue for the high density liquid phase.}
		\label{eosplots}
	\end{figure}

	Let us now consider this function plotted in Fig. \ref{eosplots} for various temperatures, having set $J_q = 1$. We clearly observe a phase transition in the (scaled) low-temperature regime where there exists a critical temperature $T_c$ below which there are three solutions while above which there is only one. This is very reminiscent of the van der Waals phase transition and indeed, as we will see in section \ref{universality}, the critical exponents do belong to the same universality class as that of van der Waals for \textit{all values} of $\gamma = J_{q/2}/J_q$. 

	\subsection{Grand potential}
	
	The grand potential is given by $\Omega=E-\mu \mathcal{Q}-T \mathcal{S}$ where $\Ss$ is the entropy density. Here we already have all the required relations except the entropy density. 
	
	The total energy $E = q^2\ex{\Hh}$ can be calculated via the effective action by considering functional derivatives. This process yields integrals over the self-energy, which we evaluate in appendix \ref{energy contri} to yield the expression in Eq. \eqref{EnergyApp} which we reproduce here for convenience
	\begin{equation}
		q\beta E = -\frac{4 K^2}{ U} \coth^{-1} \left[ \sqrt{2 \frac{K^2}{U^2}+1}\right] - 2 U\sqrt{2 \frac{K^2}{U^2} + 1}.
		\label{Energy}
	\end{equation}
	
	Motivated by the energy contributions, we consider an ansatz for $\Ss$ where the leading order is the free entropy of the system (without any interaction) and seek the leading order in $1/q$ correction. To find the leading order correction to the free entropy contribution, we use the Maxwell's relation
	\begin{equation}
		-\left(\frac{\p \Ss}{\p \Qq}\right)_{T,J_q,J_{q/2}} = \left(\frac{\p \mu}{\p T}\right)_{\Qq,J_q,J_{q/2}}
	\end{equation}
	where we already know $\mu$ using the equation of state. We find that the $\Oo(1/q)$ contribution of entropy density is at the same order of $q$ as the energy term and the explicit expression for entropy density is given by
	\begin{equation}
		\Ss = \ln 2-2 \tilde{Q}^2/q +\Oo(q^{-2})
		\label{entropy with log 2}
	\end{equation}
	We proceed to calculate the dimensionless contribution $q\beta \Omega$ which is of order $\Oo(q^0)$. We add a logarithmic contribution $q \ln 2$ in order to cancel the logarithmic contribution of the entropy in Eq. \eqref{entropy with log 2}. We denote this new expression as $\tilde{\beta} \Omegat = q \beta \Omega + q \ln 2$ where we follow the standard convention that tilded quantities are of order $\Oo(q^0)$, such as $\tilde{\beta}$ and $\Omegat$ here. We evaluate this expression to get (recall $\tilde{\Qq} \equiv \sqrt{q} \Qq = \Oo(q^0)$ in Eq. \eqref{scaling1})
	\begin{equation}
		q\beta \Omegat = q\beta E-q\beta\mu \Qq+2\tilde{\Qq}^2
	\end{equation}
	
	Next, we use the equation of state at low-temperature given in Eq. \eqref{EOS-lowT} combined with the scaling for $\beta$ and $\mu$ in eqs. \eqref{scaling2} and \eqref{scaling3}, respectively. The resulting expression for $q \beta \mu \Qq$ is of the order $\Oo(q^0)$ and is given by
	\begin{equation}
		q\beta\mu \Qq =  4\Qqt^2[1+ U\sqrt{1+2 (K/U)^2}]
	\end{equation}

	We already have the expression for $q \beta E$ in Eq. \eqref{Energy} (or Eq. \eqref{EnergyApp} as derived in appendix \ref{energy contri}) where we see that $q \beta E \sim \Oo(q^0)$. Therefore, plugging the equation of state and the energy contributions, we get
	\begin{widetext}
		\begin{equation}
			\tilde{\beta} \Omegat = -2 (1+2\Qqt^2) U\sqrt{1+2 (K/U)^2}-\frac{4 K^2}{ U} \coth^{-1} \left[ \sqrt{ [2 (K/U)^2+1]}\right] -2\tilde{\Qq}^2
		\end{equation}
	\end{widetext}
	where $q\beta \Omegat \sim \Oo(q^0)$.
	
	Next we wish to the explicit dependence of $\Omegat$ on $\qtil$ and $\ttil$. We use Eq. \eqref{U and K defs2} and recall from Eq. \eqref{def of gamma} that $\gamma \equiv J_{q/2}/J_q$ to get as the explicit expression for the grand potential
	\begin{widetext}
		\begin{equation}
			\tilde{\Omega}/J_q = -2 (1+2\tilde{\mathcal{Q}}^2) e^{-\tilde{\mathcal{Q}}^2} \sqrt{1+2 \gamma^2 e^{\tilde{\mathcal{Q}}^2}}- 4 \gamma^2 \coth^{-1} \left[ \sqrt{2 \gamma^2 e^{\tilde{\mathcal{Q}}^2}+1}\right] -2 \tilde{\mathcal{Q}}^2 \tilde{T}/J_q
			\label{OmegaFull}
		\end{equation}
	\end{widetext}
	where in the limit $\gamma \rightarrow 0$, we recover the grand potential for single large-$q$ complex SYK model (Eq. (6) of \cite{Louw2023Feb}) as it should.

	\section{Phase diagram}
	\label{universality}
	
	Having obtained the grand potential at low-temperature, we are in a position to study the phase transition as depicted in Fig. \ref{eosplots}. There exists a critical temperature $\tc$ that corresponds to the point of inflection, or in other words, when both the first derivative and the second derivative of $\mut$ with respect to $\qtil$ vanish.  
	
	To find $\tc$, we start with finding the critical charge density $\qc$ which is obtained by imposing that the second derivative of $\mut$, given in Eq. \eqref{muFull}, with respect to $\Qq$ vanish, namely $\mut^{(2)}(\qc) = 0$. This equation is satisfied trivially for $\qc = 0$ or when
	\begin{equation}
		2 \gamma ^4 e^{2 \qc^2} \left(\qc^2-3\right)+3 \gamma ^2 e^{\qc^2} \left(2 \qc^2-3\right)+2\qc^2-3=0
	\end{equation}
	Since we cannot find a simple closed form solution for $\qc$, we solve for $\gamma$ as a function of $\qc$. The positive solution is given by
	\begin{equation}\gamma^2(\qc) = e^{-\qc^2}\frac{-6  \qc^2+9- \sqrt{20 \qc^4-36 \qc^2+9}}{4 \left(\qc^2-3\right)}
	\end{equation}
	We note that $\gamma^2(\qc) \in [0, \infty)$ and therefore 
	\begin{equation}
		\qc \in \left[\sqrt{\frac{3}{2}}, \sqrt{3} \right] .
	\end{equation}
	Indeed for the case of $\gamma^2(\qc)=0$, we get $\qc = \sqrt{3/2}$ which matches with the case of single large-$q$ complex SYK model \cite{Louw2023Feb} as expected.
	
	The critical temperature is found by solving $\frac{\partial \mut}{\partial \qc} = 0$, which yields
	\begin{equation}
		\frac{\ttil_c}{J_q} = e^{-\qc^2} \frac{2 \qc^2-1+2 e^{\qc^2}\gamma^2(\qc) (\qc^2-1)}{\sqrt{1+2 e^{\qc^2}\gamma^2(\qc^2)}}
	\end{equation}
	
\begin{figure}
		\centering
		%\captionsetup{justification=centering}
		\includegraphics[width=\columnwidth]{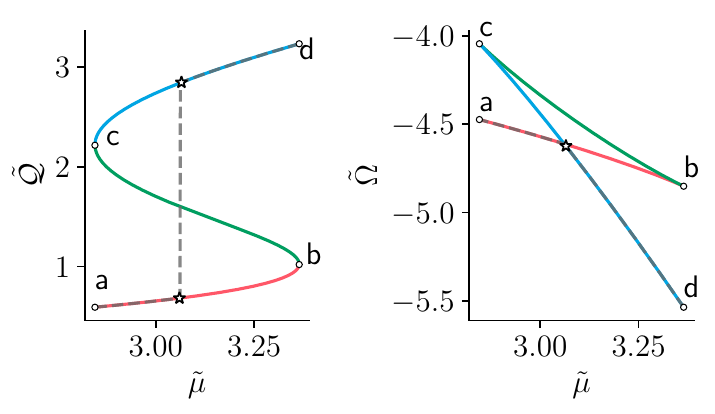}
		\caption{Plots of the equation of state (Eq. \eqref{muFull}) and grand potential (Eq. \eqref{OmegaFull}) for $\gamma^2 = 0.28$, at temperature $T=0.6 T_c$, and $J_q=1$.}
		\label{Phases}
	\end{figure}

	With this, we can now plot the parts of the EOS and grand potential corresponding to the various phases, as given in Fig. \ref{Phases}.
	
	\FloatBarrier
	
	\subsection{Universality Class}
	
	Finally, we go near the critical point on the phase diagram and expand both the chemical potential (Eq. \eqref{muFull}) and grand potential (Eq. \eqref{OmegaFull}) around the critical point, including $\muc \equiv \mut(\qc)\vert_{T = T_c}$, to get the critical exponents. To do so, we first define the reduced shifted set of variables as follows:
	\begin{equation}
		m \equiv \frac{1}{a_c}\left[\frac{\mut}{\muc}-1\right],\;\, \rho \equiv \frac{\qtil}{\qc}-1,\;\, \ttil \equiv \tc \left[1+\frac{t}{b_c}\right]
	\end{equation}
	such that the critical point corresponds to $m=\rho=t = 0$. We have further defined
	\begin{widetext}
		\begin{equation}
			\begin{aligned}
				a_c &\equiv \frac{1+4 \qc^2 (3-5\qc^2/3)+3 (2 \qc^2-3)\sqrt{(2 \qc^2/3-1)(10 \qc^2/3-1)}}{4}\\
				b_c &\equiv \frac{1-4 \qc^2 +24 (2 \qc^2/3)^2 - \sqrt{(2 \qc^2/3-1)(10 \qc^2/3-1)}}{16 (2 \qc^2/3)^2 a(\qc)}.
			\end{aligned}
		\end{equation}
	\end{widetext}
	
	We have also rescaled the reduced shifted variables such that the overlap with the solved case of single SYK model becomes more apparent. This is the origin of the constants $a(\qc),b(\qc)$.  Such a rescaling and shifting also enters into the grand potential
	\begin{equation}
		f \equiv \frac{\tilde{\Omega}-\tilde{\Omega}_c}{\muc \qc a_c} + m - t/3
	\end{equation}
	We will focus on the expansion around the critical order parameter ($\rho=0$). Using the explicit expressions in Eq. \eqref{OmegaFull} for the grand potential and Eq. \eqref{muFull} for the equation of state, we expand the reduced grand potential $f$ as well as the reduced equation of state $m$ to yield
	\begin{subequations}
		\begin{align}
			f &= -\rho^2 \frac{t + (3\rho/2)^2}{3} + \Oo(\rho^5)
			\label{critical point expansiona} \\
			m &= 2t/3 + \rho(2t/3+\rho^2) + \Oo(\rho^4)  
			\label{critical point expansionb}
		\end{align}
	\end{subequations}
	respectively. At this point, we note that we are left with the same equations as single large-$q$ complex SYK model \cite{Louw2023Feb}. As such, we also have to consider field mixing to give the correct scaling function and critical exponents \cite{Wang2007May}. By this we mean that the correct ordering field $\hh$ is not the chemical potential, but rather also gains a contribution from the temperature field, namely $\hh \equiv m - 2t/3$. We would like to express the reduced grand potential $f$ and reduced charge density $\rho$ in terms of $t$ and $\hh$. To do this, we solve for $\rho$ in the cubic equation \eqref{critical point expansionb} in terms of $t$ and $m$, use the definition of $\hh$, then substitute in Eq. \eqref{critical point expansiona} and expand in various limits of either small $t$ or small $\hh$. Since the scaling near the critical point for the reduced grand potential is found to be the same as single large-$q$ complex SYK model, the details of this procedure are already carried out in appendix C of \cite{Louw2023Feb}. For instance, around $\hh=0$ for small values of $\hh$ we get	
	\begin{equation}
		\begin{aligned}
			f(t, \hh)&=-\frac{|t|^{2}}{6}-|\hh||2 t / 3|^{1/2}-\frac{3 \hh^2}{2}|t|^{-1}+\mathcal{O}\left(\hh^3|t|^{-5 / 2}\right) \\
			\rho(t, \hh)&=\operatorname{sgn}(\hh)|2 t / 3|^{1 / 2}+2 \hh|2 t / 3|^{-1}+\mathcal{O}\left(\hh^2 t^{-5 / 2}\right)
		\end{aligned}
		\label{f around small h}
	\end{equation}
	respectively. Note that we are considering $\hh$ to be small, but not necessarily $t$ to be small such that the term inside the big-$\Oo$ notation is small. For small $t$ around $t=0$ (but not necessarily small $\hh$), we also have 
	\begin{equation}
		\begin{aligned}
			f(t, \hh)&=-\frac{3}{4}|\hh|^{4/3}\left[1+\mathcal{O}\left(t \hh^{-2 / 3}\right)\right] \\
			\rho(t, \hh)&=-\partial_{\hh} f(t, \hh)
		\end{aligned}
		\label{f around t=0}
	\end{equation}
	where $f(t, \hh)$ used in the derivative is the one that has been expanded for small $t$ in Eq. \eqref{f around t=0}. Then we have the critical exponents $\upalpha$ calculated through specific heat $C_{\hh} \propto-\partial_t^2 f(t, 0) \propto|t|^{-\upalpha}$ (using $f$ in Eq. \eqref{f around small h}), $\upbeta$ through order parameter $\rho(t, 0) \propto|t|^\upbeta$ (using $\rho$ in Eq. \eqref{f around small h}), $\upgamma$ through susceptibility $\left.\chi_\hh \propto \partial_\hh^2 f(t, \hh)\right|_{\hh=0} \propto|t|^{-\upgamma}$ (using $f$ in Eq. \eqref{f around small h}) and $\delta$ through $\rho(0, \hh)\propto \hh^{1 / \delta}$ (using $\rho$ in Eq. \eqref{f around t=0}). The critical exponents are given in table \ref{critical exponents}. These are mean-field critical exponents, and therefore our model belongs to the same universality class as van der Waals.
	
	\begin{table}
		\caption{Critical exponents}
		\begin{tabular}{c c c c}
			%		\hline\noalign{\smallskip}
			$\upalpha$ \hspace{1cm}& $\upbeta$ \hspace{1cm}& $\upgamma$ \hspace{1cm}& $\delta$\\
			\hline\noalign{\smallskip}
			0 \hspace{1cm}&$\frac{1}{2}$ \hspace{1cm}& 1  \hspace{1cm}& 3\\
			%		\hline\noalign{\smallskip}
		\end{tabular}
		\label{critical exponents}
	\end{table}
	\FloatBarrier
	
	Surprisingly, even though the expressions for the equation of state (Eq. \eqref{muFull}) and the grand potential (Eq. \eqref{OmegaFull}) are different from the case of single large-$q$ complex SYK model due to the presence of another scale $\gamma$, we get the identical scaling of both of them around the critical point as in the case of single large-$q$ SYK model. In other words, all the above critical point expansions of the reduced equation of state $m$, the reduced charge density or equivalently the order parameter $\rho$ and the reduced grand potential $f$ are the same as that for single large-$q$ complex SYK model. Therefore, our system of coupled large-$q$ SYK models belong to the same (Landau-Ginzburg mean-field) universality class as that of a single large-$q$ complex SYK model, which in turn belongs to the van der Waals (mean-field) universality class.

	\section{Dynamics and chaos}
	\label{LyapunSec}
	
	In this section, we will compute the Lyapunov exponent for our coupled system of complex large-$q$ SYK models and show that the model saturates the Maldacena-Shenker-Stanford (MSS) bound at low-temperature \cite{MSS2016}, thereby being maximally chaotic to order $\Oo{1/q}$ in the low-temperature limit.

	\subsection{4-point function}
	To obtain the Lyapunov exponent, we first analyze the four point function for the coupled system. In the large-$N$ limit, the correlator for the complex fermions $c$ is given by \cite{Maldacena2016Nov, Bhattacharya2017Nov} 
	\begin{equation}
		\begin{aligned}
			\frac{1}{N^2}\sum_{j,k}^N \langle T & c_j(\tau_1)c_j^{\dagger}(\tau_2) c_k^{\dagger}(\tau_3)c_k(\tau_4)\rangle   = \\
			& \Gg(\tau_{12})\Gg(\tau_{34}) + \frac{1}{N}   \mathcal{F}(\tau_1,\tau_2,\tau_3,\tau_4) + \dots
		\end{aligned}
		\label{otoc def 1}
	\end{equation}
	where the indices $j,k$ are the fermionic sites, and $\mathcal{F} = \sum_n \mathcal{F}_n$ is the \textit{unregularized} out-of-time-order correlator (OTOC). The first term is a disconnected diagram, and we are interested in the $\Oo{\frac{1}{N}}$ diagrams. Here, the disorder average is non-vanishing if the number of indices for the two couplings $X_{\mu_1\dots \mu_{\kappa q/2}}^{\nu_1\dots \nu_{\kappa q/2}}$ (where $\kappa = \{\frac{1}{2}, 1 \}$) are the same. As a result, the average over the product of the $q$ interaction coupling with the $q/2$ interaction is always zero. As an example, take $q=4$, for which we have the disorder average $\ex{X_{\mu_1 \mu_{2}}^{\nu_1 \nu_{2}} \;X_{\mu_1 }^{\nu_1}}=0$.
	
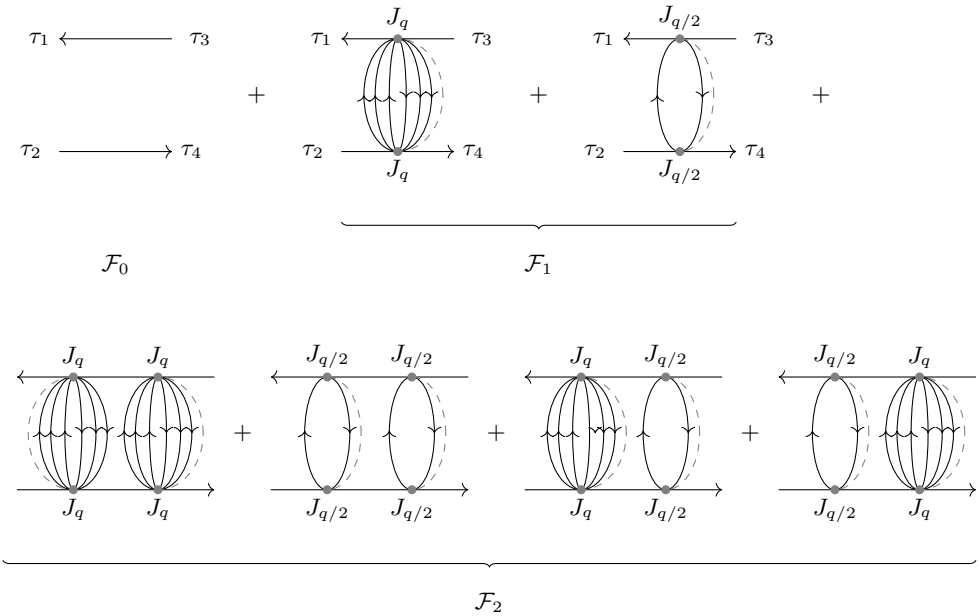
\begin{figure}[h!]
		\centering
		\begin{tikzpicture}[scale=0.75]
			\node at (2.5,4){$\tau_3$};
			\node at (-0.5,2){$\tau_2$};
			\draw[->](2,4)--(0,4) node[black, anchor=east]{$\tau_1$};
			\draw[->](0,2)--(2,2) node[black, anchor=west]{$\tau_4$};
			\node at (1,0) {$\mathcal{F}_0$};
			\node at (3.5,3){+};
			\node at (7.5,4){$\tau_3$};
			\node at (4.5,2){$\tau_2$};
			\draw[->](7,4)--(5,4) node[black, anchor=east]{$\tau_1$};
			\draw[->](5,2)--(7,2) node[black, anchor=west]{$\tau_4$};
			\draw(6,3) circle (0.4 and 1); 
			\draw(6,3) circle (0.15 and 1); 
			\draw(6,3) circle (0.6 and 1);
			\draw[->] (5.85,2.95) -- (5.85,3);
			\draw[->] (5.4,2.95) -- (5.4,3);
			\draw[->] (5.6,2.95) -- (5.6,3);
			\draw[->] (6.15,3) -- (6.15,2.95);
			\draw[->] (6.4,3) -- (6.4,2.95);
			\draw[->] (6.6,3) -- (6.6,2.95);
			\begin{scope}
				\clip (6,2) rectangle (7,4);
				\draw[gray, dashed] (6,3) circle(0.8 and 1);
			\end{scope}
			\filldraw [gray] (6,4) circle (2pt) node[black, anchor=south]{$J_q$};
			\filldraw [gray] (6,2) circle (2pt) node[black, anchor=north]{$J_q$};
			\node at (8.5,3){+};
			\node at (12.5,4){$\tau_3$};
			\node at (9.5,2){$\tau_2$};
			\draw[->](12,4)--(10,4) node[black, anchor=east]{$\tau_1$};
			\draw[->](10,2)--(12,2) node[black, anchor=west]{$\tau_4$};
			\draw(11,3) circle (0.4 and 1); 
			\draw[->] (10.6,2.95)--(10.6,3);
			\draw[->] (11.4,3)--(11.4,2.95);
			\begin{scope}
				\clip (11,2) rectangle (12,4);
				\draw[gray, dashed] (11,3) circle(0.6 and 1);
			\end{scope}
			\filldraw [gray] (11,4) circle (2pt) node[black, anchor=south]{$J_{q/2}$};
			\filldraw [gray] (11,2) circle (2pt) node[black, anchor=north]{$J_{q/2}$};  
			\node at (13.5,3){+};
			\draw [decorate,
			decoration = {brace}] (12,0.75) --  (5,0.75);
			\node at (8.5,0) {$\mathcal{F}_1$};
			
			\begin{scope}[shift={(-1,-2)}];
				%\node at (4.25,0){$t_3$};
				%\node at (-0.25,-2){$t_2$};
				\draw[->](3.75,0)--(0.25,0); %node[black, anchor=east]{$\tau_1$};
				\draw[->](0.25,-2)--(3.75,-2); %node[black, anchor=west]{$\tau_4$};
				\draw(1.25,-1) circle (0.4 and 1); 
				\draw(1.25,-1) circle (0.15 and 1); 
				\draw(1.25,-1) circle (0.6 and 1); 
				\draw(2.75,-1) circle (0.4 and 1); 
				\draw(2.75,-1) circle (0.15 and 1); 
				\draw(2.75,-1) circle (0.6 and 1);
				\draw[->] (1.4,-0.95) -- (1.4,-1);
				\draw[->] (1.65,-0.95) -- (1.65,-1);
				\draw[->] (1.85,-0.95) -- (1.85,-1);
				\draw[->] (1.1,-1) -- (1.1,-0.95);
				\draw[->] (0.85,-1) -- (0.85,-0.95);
				\draw[->] (0.65,-1) -- (0.65,-0.95);
				
				\draw[->] (2.9,-0.95) -- (2.9,-1);
				\draw[->] (3.15,-0.95) -- (3.15,-1);
				\draw[->] (3.35,-0.95) -- (3.35,-1);
				\draw[->] (2.6,-1) -- (2.6,-0.95);
				\draw[->] (2.15,-1) -- (2.15,-0.95);
				\draw[->] (2.35,-1) -- (2.35,-0.95);
				\begin{scope}
					\clip (1.25,-2) rectangle (0.25,0);
					\draw[gray, dashed] (1.25,-1) circle(0.8 and 1);
				\end{scope}
				\begin{scope}
					\clip (2.75,-2) rectangle (3.75,0);
					\draw[gray,dashed](2.75,-1) circle (0.8 and 1);
				\end{scope}
				\filldraw [gray] (1.25,0) circle (2pt) node[black, anchor=south]{$J_q$};
				\filldraw [gray] (1.25,-2) circle (2pt) node[black, anchor=north]{$J_q$};
				\filldraw [gray] (2.75,0) circle (2pt) node[black, anchor=south]{$J_{q}$};
				\filldraw [gray] (2.75,-2) circle (2pt) node[black, anchor=north]{$J_{q}$}; 
				\node at (4.25,-1){+};
				%\node at (10.25,0){$t_3$};
				%\node at (5.75,-2){$t_2$};
				\draw[->](8.25,0)--(4.75,0); %node[black, anchor=east]{$t_1$};
				\draw[->](4.75,-2)--(8.25,-2); %node[black, anchor=west]{$t_4$};
				\draw(5.75,-1) circle (0.4 and 1); 
				\draw(7.25,-1) circle (0.4 and 1);
				\draw[->] (5.35,-1) -- (5.35, -0.95);
				\draw[->] (6.15,-0.95) -- (6.15,-1);
				\draw[->] (6.85,-1) -- (6.85,-0.95);
				\draw[->] (7.65,-0.95) -- (7.65, -1);
				
				\begin{scope}
					\clip (5.75,-2) rectangle (6.75,0);
					\draw[gray, dashed] (5.75,-1) circle(0.6 and 1);
				\end{scope}
				\begin{scope}
					\clip (7.25,-2) rectangle (8.25,0);
					\draw[gray,dashed](7.25,-1) circle (0.6 and 1);
				\end{scope}
				\filldraw [gray] (5.75,0) circle (2pt) node[black, anchor=south]{$J_{q/2}$};
				\filldraw [gray] (5.75,-2) circle (2pt) node[black, anchor=north]{$J_{q/2}$};
				\filldraw [gray] (7.25,0) circle (2pt) node[black, anchor=south]{$J_{q/2}$};
				\filldraw [gray] (7.25,-2) circle (2pt) node[black, anchor=north]{$J_{q/2}$}; 
				\node at (8.75,-1){+};
				%\node at (16.25,0){$t_3$};
				%\node at (11.75,-2){$t_2$};
				\draw[->](12.75,0)--(9.25,0); %node[black, anchor=east]{$t_1$};
				\draw[->](9.25,-2)--(12.75,-2); %node[black, anchor=west]{$t_4$};
				\draw(10.25,-1) circle (0.4 and 1); 
				\draw(10.25,-1) circle (0.15 and 1); 
				\draw(10.25,-1) circle (0.6 and 1); 
				\draw(11.75,-1) circle (0.4 and 1); 
				\draw[->] (9.65,-1) -- (9.65,-0.95);
				\draw[->] (9.85,-1) -- (9.85,-0.95);
				\draw[->] (10.1,-1) -- (10.1,-0.95);
				\draw[->] (10.65,-0.95) -- (10.65,-1);
				\draw[->] (10.85,-0.95) -- (10.85,-1);
				\draw[->] (10.5,-0.95) -- (10.5,-1);
				\draw[->] (11.35,-1) -- (11.35, -0.95);
				\draw[->] (12.15,-0.95) -- (12.15,-1);
				\begin{scope}
					\clip (10.25,-2) rectangle (11.25,0);
					\draw[gray,dashed](10.25,-1) circle (0.8 and 1);
				\end{scope}
				\begin{scope}
					\clip (11.75,-2) rectangle (12.75,0);
					\draw[gray,dashed](11.75,-1) circle (0.6 and 1);
				\end{scope}
				\filldraw [gray] (10.25,0) circle (2pt) node[black, anchor=south]{$J_q$};
				\filldraw [gray] (10.25,-2) circle (2pt) node[black, anchor=north]{$J_q$};
				\filldraw [gray] (11.75,0) circle (2pt) node[black, anchor=south]{$J_{q/2}$};
				\filldraw [gray] (11.75,-2) circle (2pt) node[black, anchor=north]{$J_{q/2}$}; 
				\node at (13.25,-1){+};
				\draw[->](17.25,0)--(13.75,0); %node[black, anchor=east]{$t_1$};
				\draw[->](13.75,-2)--(17.25,-2); %node[black, anchor=west]{$t_4$};
				\draw(16.25,-1) circle (0.4 and 1); 
				\draw(16.25,-1) circle (0.15 and 1); 
				\draw(16.25,-1) circle (0.6 and 1); 
				\draw(14.75,-1) circle (0.4 and 1); 
				\draw[->] (14.35,-1) -- (14.35, -0.95);
				\draw[->] (15.15,-0.95) -- (15.15,-1);
				
				\draw[->] (16.1,-1) -- (16.1, -0.95);
				\draw[->] (16.4,-0.95) -- (16.4,-1);
				\draw[->] (15.65,-1) -- (15.65, -0.95);
				\draw[->] (16.85,-0.95) -- (16.85,-1);
				\draw[->] (15.85,-1) -- (15.85, -0.95);
				\draw[->] (16.65,-0.95) -- (16.65,-1);
				\begin{scope}
					\clip (16.25,-2) rectangle (17.25,0);
					\draw[gray,dashed](16.25,-1) circle (0.8 and 1);
				\end{scope}
				\begin{scope}
					\clip (14.75,-2) rectangle (15.75,0);
					\draw[gray,dashed](14.75,-1) circle (0.6 and 1);
				\end{scope}
				\filldraw [gray] (16.25,0) circle (2pt) node[black, anchor=south]{$J_q$};
				\filldraw [gray] (16.25,-2) circle (2pt) node[black, anchor=north]{$J_q$};
				\filldraw [gray] (14.75,0) circle (2pt) node[black, anchor=south]{$J_{q/2}$};
				\filldraw [gray] (14.75,-2) circle (2pt) node[black, anchor=north]{$J_{q/2}$}; 
				\draw [decorate,
				decoration = {brace}] (17.25,-3.25) --  (0,-3.25);
				\node at (8.62,-4) {$\mathcal{F}_2$};
			\end{scope}
		\end{tikzpicture}
		\caption{The diagrams related to the $\Oo{\frac{1}{N}}$ part of the correlator for our system of coupled complex large-$q$ SYK models. The diagram is drawn for $q=8$, and also consists of a diagram with $\tau_3 \leftrightarrow \tau_4$, which have been omitted here. There are $q-2$ lines connecting the two horizontal rails, out of which half of them ($q/2 - 1$ lines) run in one direction while the remaining half ($q/2 - 1$ lines) run in the opposite.}
		\label{4pointdiag}
	\end{figure}

	Unlike the single large-$q$ complex SYK model, where any particular $\mathcal{F}_n$ in the sum denotes a single diagram \cite{Bhattacharya2017Nov}, in our model each $\mathcal{F}_n$ represents a sum of ladder diagrams as depicted in Fig. (\ref{4pointdiag}). This is after resumming the perturbative terms to get the ladder diagrams in terms of the dressed propagators \cite{Klebanov2017Feb}.
	
	\onecolumngrid

	\FloatBarrier
	\twocolumngrid
	
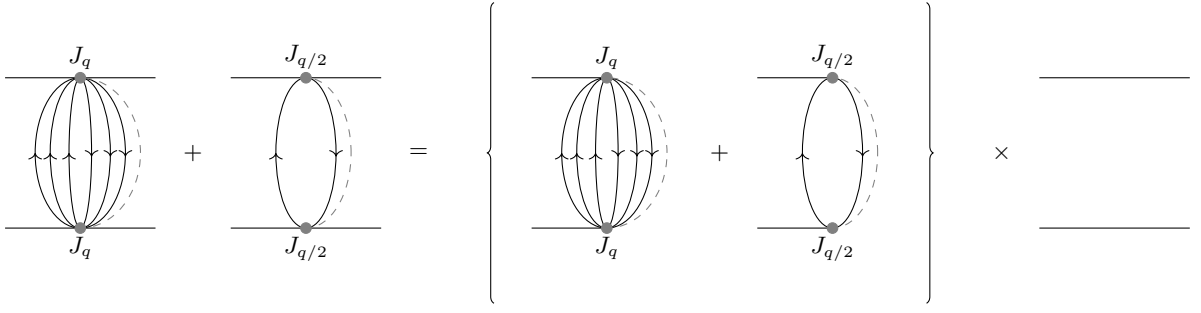
\begin{figure}[h!]
		\centering
		\begin{tikzpicture}[scale=1.0]
			%\node at (7.5,4){$t_3$};
			%\node at (4.5,2){$t_2$};
			\draw[](2,4)--(0,4); %node[black, anchor=east]{$t_1$};
			\draw[](0,2)--(2,2); %node[black, anchor=west]{$t_4$};
			\draw(1,3) circle (0.4 and 1); 
			\draw(1,3) circle (0.15 and 1); 
			\draw(1,3) circle (0.6 and 1); 
			\draw[->] (0.6, 2.95)--(0.6,3);
			\draw[->] (0.4, 2.95)--(0.4,3);
			\draw[->] (0.85, 2.95)--(0.85,3);
			\draw[->] (1.6, 3)--(1.6,2.95);
			\draw[->] (1.4, 3)--(1.4,2.95);
			\draw[->] (1.15, 3)--(1.15,2.95);
			\begin{scope}
				\clip (1,2) rectangle (2,4);
				\draw[gray,dashed](1,3) circle (0.8 and 1);
			\end{scope}
			\begin{scope}
				\clip (4,2) rectangle (5,4);
				\draw[gray,dashed](4,3) circle (0.6 and 1);
			\end{scope}
			\filldraw [gray] (1,4) circle (2pt) node[black, anchor=south]{$J_q$};
			\filldraw [gray] (1,2) circle (2pt) node[black, anchor=north]{$J_q$};
			\node at (2.5,3){+};
			%\node at (12.5,4){$t_3$};
			%\node at (9.5,2){$t_2$};
			\draw[](5,4)--(3,4); %node[black, anchor=east]{$t_1$};
			\draw[](3,2)--(5,2); %node[black, anchor=west]{$t_4$};
			\draw(4,3) circle (0.4 and 1); 
			\draw[->] (3.6,2.95)--(3.6,3);
			\draw[->] (4.4,3)--(4.4,2.95);
			\filldraw [gray] (4,4) circle (2pt) node[black, anchor=south]{$J_{q/2}$};
			\filldraw [gray] (4,2) circle (2pt) node[black, anchor=north]{$J_{q/2}$};
			
			\node at (5.5,3){=};
			\begin{scope}[shift={(0.5,0)}];
				\draw [decorate,
				decoration = {brace}] (6,1) --  (6,5);
				\draw[](6.5,4)--(7.5,4); %node[black, anchor=east]{$t_1$};
				\draw[](7.5,2)--(6.5,2);
				\draw(7.5,3) circle (0.4 and 1); 
				\draw(7.5,3) circle (0.15 and 1); 
				\draw(7.5,3) circle (0.6 and 1); 
				
				\draw[->] (7.9,3)--(7.9, 2.95);
				\draw[->] (7.65, 3)--(7.65, 2.95);
				\draw[->] (8.1, 3)--(8.1, 2.95);
				
				\draw[->] (7.1,  2.95)--(7.1,3);
				\draw[->] (6.9,  2.95)--(6.9,3);
				\draw[->] (7.35,  2.95)--(7.35,3);
				\begin{scope}
					\clip (10.5,2) rectangle (11.5,4);
					\draw[gray,dashed](10.5,3) circle (0.6 and 1);
				\end{scope}
				\begin{scope}
					\clip (7.5,2) rectangle (8.5,4);
					\draw[gray,dashed](7.5,3) circle (0.8 and 1);
				\end{scope}
				\filldraw [gray] (7.5,4) circle (2pt) node[black, anchor=south]{$J_q$};
				\filldraw [gray] (7.5,2) circle (2pt) node[black, anchor=north]{$J_q$};
				
				%\node at (7.25,-3){$t_2$};
				\node at (9,3){+};
				\draw[](10.5,4)--(9.5,4); %node[black, anchor=east]{$t_1$};
				\draw[](9.5,2)--(10.5,2);
				\draw(10.5,3) circle (0.4 and 1); 
				\draw[->] (10.1,2.95)--(10.1,3);
				\draw[->] (10.9,3)--(10.9,2.95);
				\filldraw [gray] (10.5,4) circle (2pt) node[black, anchor=south]{$J_{q/2}$};
				\filldraw [gray] (10.5,2) circle (2pt) node[black, anchor=north]{$J_{q/2}$};
				\draw [decorate,
				decoration = {brace}] (11.75,5) --  (11.75,1);
				%\node at (10.25,-3){$t_2$};
				\node at (12.75,3){$\times$};
				%\node at (15.25,-1){$t_3$};
				\draw[](15.25,4)--(13.25,4) ;
				\draw[](13.25,2)--(15.25,2); %node[black, anchor=west]{$t_4$};
			\end{scope}
			
			%\draw [decorate,
			%  decoration = {brace}] (5,0) --  (0,0);
			% \draw [decorate,
			% decoration = {brace}] (11.5,0) --  (7,0);
			%\node at (2.5,-1){$\mathcal{F}_1$};
			%\node at (6,-1){=};
			%\node at (9.25,-1){$K$};
			%\node at (12.75,-1){$\times$};
			% \node at (14.75,-1){$\mathcal{F}_0$};
		\end{tikzpicture}
		\caption{In the large-$N$ limit, the diagrams of order $\Oo{\frac{1}{N}}$ in the out-of-time-correlator (OTOC) for the coupled SYK system can be constructed with a kernel $K$. The OTOC is then written as $\mathcal{F} = \mathcal{F}_0 + K \mathcal{F}$, where $\mathcal{F} = \sum\limits_{n=0}^{\infty}\mathcal{F}_n$, where $\mathcal{F}_0$ is defined in Fig. \ref{4pointdiag} and becomes suppressed at large time due to exponential growth of the OTOC. This figure shows the process $\mathcal{F}_1 = K \mathcal{F}_0$ for $q=8$, where the term in parentheses is the kernel.}
		\label{fig:kernel}
	\end{figure}

	According to standard techniques \cite{Kitaev2015, Bhattacharya2017Nov, Maldacena2016Nov}, each diagram $\mathcal{F}_{n+1}$ is generated by applying a kernel $K$ to the previous diagram $\mathcal{F}_n$, i.e., 
	\begin{equation}
		\mathcal{F}_{n+1}(\tau_1,\tau_2,\tau_3,\tau_4) = \int d\tau d \tau' K(\tau_1,\tau_2,\tau,\tau') \mathcal{F}_n(\tau,\tau',\tau_3,\tau_4)
		\label{F integral}
	\end{equation}
	or $\mathcal{F}_{n+1} = K \odot \mathcal{F}_n$ in short. We can recursively repeat the process and assume asymptotic exponential growth of the OTOC. The zero rung diagrams are also negligible at large times, which gives an eigenvalue problem of type $\mathcal{F} = K \odot F$ (see below for more details). The kernel for the coupled system at equilibrium is given by 
	\begin{widetext}
		\begin{equation}
			K (\tau_1, \tau_2; \tau, \tau')\equiv - \sum_{\kappa = \{ \frac{1}{2}, 1\}} \frac{2^{\kappa q-1}J_{\kappa q}^2 }{ q} (\kappa q-1)\; \Gg(\tau_1-\tau) \Gg(\tau_{2}-\tau') \Gg^>(\tau-\tau')^{\kappa q/2-1}  \Gg^<(\tau'-\tau)^{\kappa q/2-1}
			\label{K def 1}
		\end{equation}
	\end{widetext}
	and the generation of the ladder diagrams with the kernel is illustrated in Fig. (\ref{fig:kernel}). A notable result is that our kernel does not merely add a rung to the ladder diagram. It adds a sum of diagrams, where each diagram is extended with one rung. 
	
	\onecolumngrid
	
	\FloatBarrier
	\twocolumngrid

	\subsection{Chaos regime}
	
	We wish to obtain the exponential growth rate of the OTOC at late times. For this, we utilize the standard approach \cite{Kitaev2015, MSS2016} by considering the following \textit{regularized} out-of-time-order correlator (OTOC) in real time as
	\begin{equation}
		\mathcal{F}(t_1,t_2, t_3, t_4) = \Tr\left[y \;c_j(t_1)\, y \, c_k^{\dagger}(t_3) \,y \,c_j^{\dagger}(t_2)\, y\, c_k(t_4)\right],
		\label{otoc def 2}
	\end{equation}
	having defined $y\equiv \rho(\beta)^{1/4}$, where $\rho(\beta)$ is the thermal density matrix. In this approach, the fermions are separated by a quarter of the thermal circle, and a real-time separation between the pairs of fermions. More on the specifics of this approach can be found in appendix \ref{app:chaos}. We notice that the kernel generates all the infinite diagrams, with the exception of the free propagator, i.e., 
	\begin{equation}
		\sum\limits_{n=1}^{\infty} \mathcal{F}_n = K \odot \sum\limits_{n=0}^{\infty} \mathcal{F}_n.
	\end{equation}
	
	Adding $\mathcal{F}_0$ on both sides and assuming asymptotic exponential growth rate of OTOC, the zero rung diagram $\mathcal{F}_0$ gets suppressed in the limit $t_1, t_2 \to \infty$. We get an eigenvalue equation of the type \cite{Bhattacharya2017Nov, Maldacena2016Nov} $\mathcal{F} = K \odot \mathcal{F}$, namely
	\begin{equation}
		\mathcal{F}(t_1,t_2, t_3, t_4) = \nint[-\infty][\infty]{t}\nint[-\infty][\infty]{t'} K_R(t_{1},t_2;t, t')  \mathcal{F}(t, t', t_3,t_4).
		\label{f-k integral}
	\end{equation}
	
	The exponential growth of the OTOC is determined by the real-time part of $\mathcal{F}$ , and the diagrams describing the OTOC are generated by the \textit{retarded kernel}. Given the time translational invariance for our equilibrium setup, we have (with the notation from section \ref{model section}, namely $t_{ij} \equiv t_i - t_j$)
	\begin{widetext}
		\begin{equation}
			K_R(t_1, t_2 ; t_3, t_4) =  \sum_{\kappa = \{\frac{1}{2},1\}}\frac{2^{\kappa q-1}J_{\kappa q}^2}{q}  (\kappa q-1)\Gg_R(t_{13}) \Gg_R(-t_{24})  \Gg_{lr}^>(t_{34})^{\kappa q/2-1} \Gg_{lr}^<(-t_{34})^{\kappa q/2-1}, \label{KR def}
		\end{equation}
	\end{widetext}
	where $\Gg_R$ and $\Gg_{lr}$ are the retarded propagator and the Wightman correlator respectively, defined as 
	\begin{equation}
		\begin{aligned}
			\Gg_R (t) &\equiv [\Gg^{>}(t)-\Gg^{<}(t)]\; \theta(t)\stackrel{q\rightarrow \infty}{=} \theta(t), \\
			\Gg^{>}_{lr}( t) &\equiv  \Gg^{>}( \imath t + \beta/2), \\
			\Gg^{<}_{lr}( t) &\equiv  \Gg^{<}( \imath t - \beta/2).
		\end{aligned}
		\label{retard and wightman}
	\end{equation}
	Note the relation in frequency domain between the Matsubara Green's function and real time retarded Green's function, namely $\Gg(\i \omega_n \to \omega + \i 0^+) = \Gg_R(\omega)$ \cite{Jishi2013Apr}. Compared with the kernel in Eq. \eqref{K def 1}, we see that the overall minus sign vanishes. This is due to the transformation  $\tau \to i t$ and $\tau' \to i t'$, leading to $d\tau d\tau ' = -dt dt'$  in the integral of type of Eq. \eqref{f-k integral}, where the minus sign is absorbed in the retarded kernel. For a detailed discussion on the retarded kernel and the chaos exponents in real time, see \cite{Murugan2017Aug} (in particular, section 8).
	
	The explicit form of the Wightman function can be found with the Green's function in eqs. (\ref{GreenDef1}) and (\ref{solution for g}). Combined with the effective coupling (\ref{effective coupling0}), we have (recall $\cosh(-t) = \cosh(t)$)
	
	\begin{equation}
		K_R(t_{1},t_2;t_{3},t_4) =  \theta(t_{13}) \theta(t_{24}) B(t_{34}),
		\label{KR def 2}
	\end{equation}
	where
	\begin{widetext}
		\begin{equation}
			B(t_{34}) \equiv  \frac{2 (\pi v)^2 A^2 \;  \theta(t_{13}) \theta(t_{24})}{\beta^2 \left[1+\sqrt{A^2+1}\cosh(\frac{\pi v t_{34}}{\beta})\right]^2}       + \frac{ (\pi v)^2  \;  \theta(t_{13}) \theta(t_{24})}{\beta^2 \left[1+\sqrt{A^2+1}\cosh(\frac{\pi v t_{34}}{\beta})\right]}.
		\end{equation}
	\end{widetext}
	
	Recall the definition of $A$ from Eq. \eqref{A}, namely $A \equiv \frac{\pi v \beta \mathcal{J}_q}{\left(\beta \mathcal{J}_{q / 2}\right)^2}$ $= \frac{\pi v U}{q K^2}$ where we used definitions in Eq. \eqref{U and K defs} and the scaling for $\beta$ in Eq. \eqref{scaling2}. It is important to notice that even if the kernel applies multiple diagrams, the total sum of infinite diagrams in $\mathcal{F}$  does not change with applying the retarded kernel. Furthermore, the zero rung ladder diagram gets suppressed at larger times. This implies that in the coupled SYK model, $\mathcal{F}$ is also an eigenfunction of the integral transform that can be further simplified by taking the two derivatives $\partial_{t_1} \partial_{t_2}$ on both sides of the integral equation (Eq. \eqref{f-k integral}) to get
	\begin{widetext}
		\begin{equation}
			\partial_{t_1} \partial_{t_2} \mathcal{F}(t_1,t_2, t_3, t_4) =  \int_{-\infty}^{\infty} dt  \int_{-\infty}^{\infty} dt' \; \delta(t_1 - t) \delta(t_2 - t') B(t_{34})  \;\mathcal{F}(t, t', t_3,t_4).
			\label{otoc integral}
		\end{equation}
	\end{widetext}
	The delta functions enforces $t = t_1$ and $t' = t_2$. Due to the theta function and our interest in large times $t_1$ and $t_2$, we can put $t_3 = t_4 = 0$ without loss of generality. For brevity, we write $\mathcal{F}(t_1, t_2, t_3=0, t_4=0) = \mathcal{F}(t_1, t_2)$.

	\subsection{The Lyapunov exponent}

\begin{figure}[ht]
		\centering
		\includegraphics[width=0.75\columnwidth]{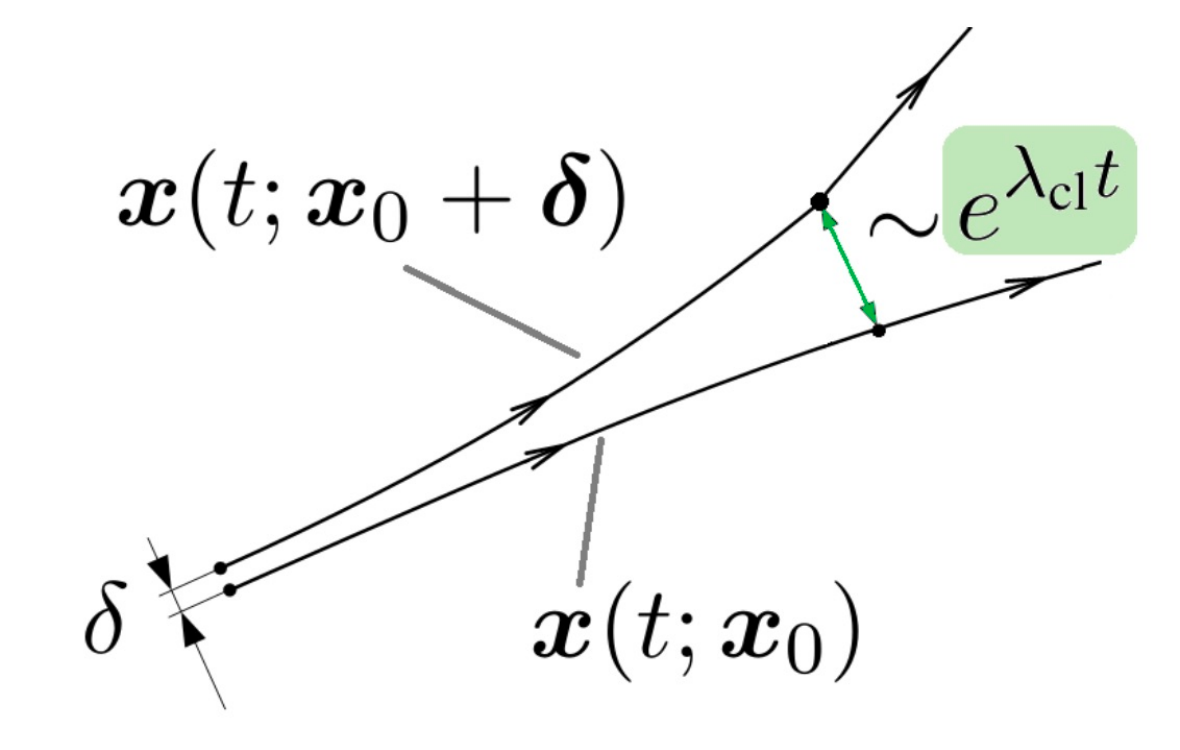}
		\caption{Classical chaos is defined by exponential sensitivity to initial conditions: classical trajectories when separated by infinitesimally small distance diverge exponentially with time which defines the classical Lyapunov exponent $\lambda_{\text{cl}}$. The Lyapunov exponent is taken as a measure of classical chaos.}
		\label{fig:lyapunov}
	\end{figure}

	Classically, the Lyapunov exponent measures sensitivity to initial conditions. Mathematically, consider two classical trajectories $\vec{x}(t, \vec{x}_0)$ and $\vec{x}(t, \vec{x}_0 + \vec{\delta})$, where $\vec{\delta}$ is infinitesimally small (see Fig. \ref{fig:lyapunov}). The classical Lyapunov exponent $\lambda_{\text{cl}}$ is defined as:
	\begin{equation}
		\left| \frac{\partial x^i(t)}{\partial x^j(0)} \right| = \left| \{x^i(t), p^j(0)\}_{\text{Poisson bracket}} \right| \sim e^{\lambda_{\text{cl}} t}
	\end{equation}
	where $x^i(t)$ is the $i^{\text{th}}$ component of $\vec{x}(t)$ and $p^j(0)$ is the $j^{\text{th}}$ component of conjugate momentum $\vec{p}$ corresponding to $\vec{x}(0)$. If $\lambda_{\text{cl}}>0,$ this indicates that two trajectories, which start from infinitesimally close initial conditions, still diverge exponentially over time.
	
	To generalize this to the quantum case, we use the standard prescription of replacing the Poisson bracket with the commutator where we elevate position $\vec{x}$ and momentum $\vec{p}$ to operators $\hat{X}$ and $\hat{P}$ respectively. In other words, we have
	\begin{equation}
		\begin{aligned}
			\left| \{x^i(t), p^j(0)\}_{\text{Poisson bracket}} \right|^2 \to & \left\langle \left| \left[ \hat{X}(t), \hat{P}(0) \right] \right|^2 \right\rangle \\
			& = \text{TOC} - 2 \Re \left[ \text{OTOC} \right]
		\end{aligned}
	\end{equation}
	where we have suppressed the proportionality factor $\frac{1}{\i \hbar}$ in front of the commutator for notational convenience. Here $\left\langle \ldots \right\rangle$ denotes the expectation value corresponding to the equilibrium Hamiltonian $\mathcal{H}$ or the associated density matrix $\rho$. TOC is the time-ordered correlator given by:
	\begin{equation}
		\text{TOC} = || \rho^{1/2} \hat{X}(t) \hat{P}(0) ||^2_F + || \hat{X}(t) \hat{P}(0) \rho^{1/2} ||^2_F
	\end{equation}
	where $||\ldots||$ denotes the Frobenius norm defined by $||A||_F \equiv \sqrt{\text{Tr}[A^\dagger A]}$. OTOC is the out-of-time-ordered correlator given by:
	\begin{equation}
		\text{OTOC} = \text{Tr} \left[ \left( \hat{X}(t) \hat{P}(0) \right)^2 \rho \right] = \left\langle \left( \hat{X}(t) \hat{P}(0) \right)^2 \right\rangle
	\end{equation}
	Note that we used position and conjugate momentum as representative examples for defining OTOC. This definition can be used for any non-commuting quantum observable. This approach is used to define and expand OTOC for large-$N$ theories in Eq. \eqref{otoc def 1}, using the regularized method for calculating OTOC introduced in \cite{MSS2016}, as employed in Eq. \eqref{otoc def 2} (also see Fig. 2 in \cite{Roberts2017Apr}). Assuming there is an exponential divergence of OTOC as $e^{\lambda_L t}$, this generalizes the classical Lyapunov exponent $\lambda_{\text{cl}}$ to the quantum Lyapunov exponent $\lambda_L$, which we now evaluate using the integral in Eq. \eqref{otoc integral}.

	To solve this integral in real time, assuming a chaotic behavior at asymptotic times, we make the standard ansatz \cite{Kitaev2015, Maldacena2016Nov, Bhattacharya2017Nov} 
	\begin{equation}
		\mathcal{F}(t_1,t_2) = e^{\lambda_L(t_1+t_2)/2} f(t_{12}),
		\label{ansatz}
	\end{equation}
	where $\lambda_L$ is the growth exponent. With the ansatz (\ref{ansatz}), and assuming the function $\mathcal{F}(t_1,t_2)$ decays at the boundaries $t_1,t_2 \to -\infty$, this reduces to a second order differential equation
	\begin{equation}
		\left[\partial_{t_1} \partial_{t_2} - B(t_{12})\right] f(t_{12}) = -\frac{\lambda_L^2}{4} f(t_{12}).
		\label{schro0}
	\end{equation}
	
	In fact, it is a ``Schr\"{o}dinger equation'' with
	a potential 
	\begin{equation}
		V(x) \equiv - 2 B(t_{12})/(\pi v T)^2,\qquad x\equiv \pi v T t_{12}/2.
	\end{equation}
	From the chain rule we then have
	$$\partial_{t_1} \partial_{t_2} = \frac{\partial x}{\p t_1} \frac{\partial x}{\p t_2} \p_x^2 = -\frac{(\pi v T)^2}{2} \p_x^2/2$$
	which leaves us with
	\begin{equation}
		\left[-\p_x^2/2 +V(x)\right] f(t_{12}) = -\frac{\lambda_L^2}{2 (\pi v T)^2} f(t_{12})\label{schro1}.
	\end{equation}
	While there typically exists an entire spectrum of Lyapunov exponents, the system's chaos is defined by the maximum $\lambda_L$. This corresponds to the minimal negative eigenvalue, which would correspond to the ground state energy $E_0$.  As such, we have a ground state wave function $\psi_0(x) \equiv f(t_{12})$ and energy 
	\begin{equation}
		E_0 \equiv -\frac{\lambda_L^2}{2 [\pi v T]^2}. \label{energyLyap}
	\end{equation}
	This leaves us with the more familiar form of the Schr\"odinger equation 
	\begin{equation}
		\left[-\frac{\p_x^2}{2} +V(x)\right] \psi_0(x) = E_0 \psi_0(x).
		\label{schro}
	\end{equation}

	\subsubsection{Perturbation theory}
	
	Let us now consider the above Schr\"odinger type equation \eqref{schro} and in particular the potential, which may be written as
	\begin{equation}
		\begin{aligned}
			V(x) %=& \frac{-4 A^2}{[ 1 + \sqrt{A^2 + 1} \cosh(2 x)]^2}\\
			%&+ \frac{-2}{[1 + \sqrt{A^2 + 1} \cosh(2 x)]}\\
			=& \frac{1}{(\sqrt{A^2 + 1}-1)/2 + \sqrt{A^2 + 1} \,/V_0(x)}\\
			&-\frac{A^2}{[(\sqrt{A^2 + 1}-1)/2 + \sqrt{A^2 + 1} \,/V_0(x)]^2}
		\end{aligned}
		\label{potential for lyapunov}
	\end{equation}
	where $A \equiv \frac{\pi v \beta \mathcal{J}_q}{\left(\beta \mathcal{J}_{q / 2}\right)^2}$ $= \frac{\pi v U}{q K^2}$. We now consider this for small $A$, with corrections $\psi_0 = \psi_0^{(0)} + A^{2} \psi_0^{(1)} + \ldots$ and $E_0 = E_0^{(0)} + A^{2} E_0^{(1)} + \ldots$ obtained via perturbation theory.
	
	For $A=0$, we are left with the P\"{o}schl-Teller potential $V_0(x) =-\text{sech}^2(x)$ \cite{Poschl1933Mar} and the ground state is 
	\begin{equation}
		\psi_0^{(0)}(x) = \text{sech}(x)/\sqrt{2}. 
	\end{equation}
	with corresponding eigenvalue is $E_0^{(0)} = -1/2$. The leading order perturbed potential is
	\begin{equation}
		V(x) = V_0(x)- \frac{V_0(x)}{4}(2+33 V_0(x))A^2 + \Oo(A^4)
	\end{equation}
	As such, first order perturbation theory yields the energy correction $E_0 = -1/2 + A^2 E^{(1)}_0+ \Oo(A^4)$, with 
	\begin{equation}
		E^{(1)}_0 = - \langle \psi^{(0)}_0|\frac{\hat{V}_0}{4}(2+33 \hat{V}_0)|\psi^{(0)}_0 \rangle.
	\end{equation}
	Inserting the identity $\int dx |x \rangle \langle x | = 1$ yields
	\begin{equation}
		\begin{aligned}
			E^{(1)}_0 %&= - \nint{x} \psi^{(0)}(x)^2\frac{V_0(x)}{4}(2+33 V_0(x))\\
			&= \nint{x} \frac{V_0^2(x)}{8}(2+33 V_0(x))
		\end{aligned}
	\end{equation}
	This integrates to $E^{(1)}_0 = -61/15$, hence the energy is lowered. Since we have the Lyapunov exponent \eqref{energyLyap}
	\begin{equation}
		\lambda_L = \sqrt{2|E_0|} \pi v T,
	\end{equation}
	one might assume that this leads to an increase to the exponents, hence allowing one to surpass the MSS bound. Note however that this correction is actually at quadratic order since
	\begin{equation}
		A = \frac{\pi v U}{q K^2}
	\end{equation}
	whereas there is another correction to $v$ in Eq. \eqref{vSol} which reduces the Lyapunov exponent at linear order, namely
	\begin{equation}
		v=2 - \frac{4}{q K} \sqrt{2 + (U/K)^2} + \Oo(q^{-2}) 
	\end{equation}
	As such the MSS bound $\lambda_L \to 2\pi T$ remains intact in the low $T$ limit. We can now return to the topic of $v$ being twice that of the $q$-body SYK model. The point is that $v$ is not necessarily a direct measure of chaos in the usual sense of single SYK model. While $\lambda_L = \sqrt{2|E_0|} \pi v T$ remains true, if we had instead first set $\gamma=0$ (implying $A \to \infty$), we would have found a potential using Eq. \eqref{potential for lyapunov}
	\begin{equation}
		V(x) \sim  -\frac{1}{[1/2 + 1/V_0(x)]^2} = V_0(2x)/4.
	\end{equation}
	This then leads to a ground state energy of $E_0 = 2$ and $v\to 1$ in the low-temperature regime \cite{Maldacena2016Nov}. Said another way, while $v$ might saturate to either $1$ or $2$, in either of these cases, the Lyapunov exponent will saturate to the MSS bound.

	\section{Conclusion and Outlook}
	\label{conclusion section}
	\subsection{Conclusion}
	Motivated from the study of a single large-$q$ complex SYK model, we considered a system of coupled SYK models in this work and studied its thermodynamic and dynamic properties. Despite having a new scale in the system, namely the ratio of interaction strengths of the two SYK terms in the Hamiltonian, we found that there exists a continuous phase transition at low temperatures just like the single SYK and the associated critical exponents correspond to the same universality class as Landau-Ginzburg (mean-field) exponents, also shared by van der Waals gas. Coincidentally, various AdS black holes also belong to the same universality class \cite{Kubiznak2012Jul, Majhi2017Oct}. We will comment more on this below. Finally, we calculated the Lyapunov exponent, and we again found that the coupled SYK system  also saturates the Maldacena-Shenker-Stanford (MSS) bound in large-$q$ limit at low-temperature, thereby making our model maximally chaotic, surprisingly matching the behavior of the single SYK model.
	
	We now proceed to discuss the ``very'' low-temperature regime, where we can make connections with the Hawking-Page phase transition on the holographically dual black holes. We present preliminary calculations and clearly highlight the steps that can be taken as future work to make the connection robust with the first-order Hawking-Page transition, if at all possible. 
	
	\subsection{Relation to the phase transitions at ``very'' low-temperature}
	\label{very low temperatures}
	
	Given that the MSS bound is always saturated over the phase transition corresponding to the temperature regime $T = \Oo(q^{-1})$ \eqref{scaling2}, we conclude that both liquid and gaseous phases are maximally chaotic. In other words, we have a maximally chaotic to maximally chaotic phase transition similar to those found in small to large black hole phase transitions \cite{Louw2023Oct,Kubiznak2012Jul}.
	
\begin{figure}[ht]
		\centering
		\includegraphics[width=\columnwidth]{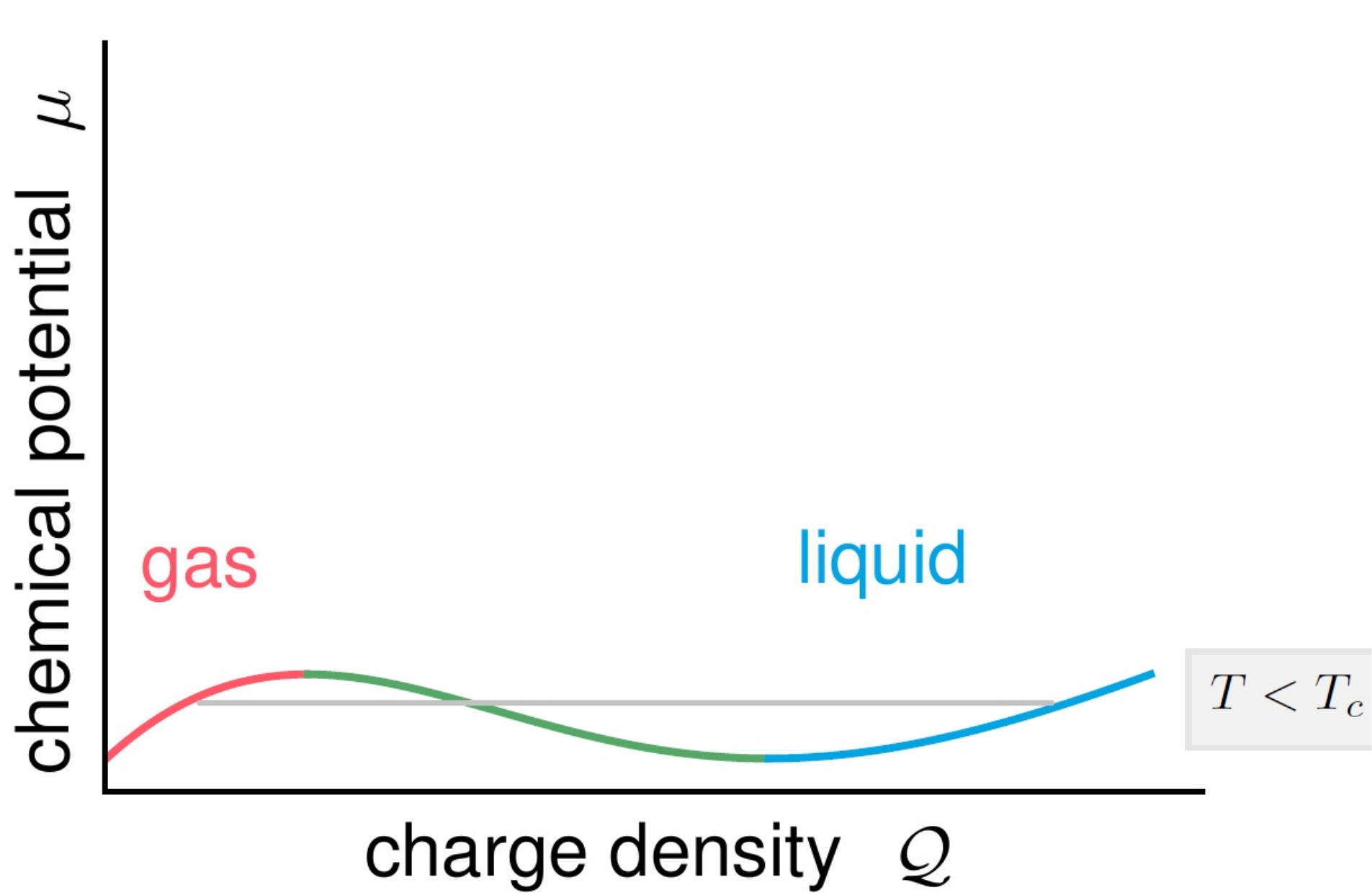}
		\caption{Liquid and gaseous phases for temperatures below the critical temperature, where we have borrowed terminology from the van der Waals phase transition that belongs to the same universality class as our coupled SYK system. Color coding is the same as in Figs. \ref{eosplots} and \ref{Phases}.}
		\label{liquid and gas phase}
	\end{figure}

	This, however, changes in the rescaled regime of \textit{very} low-temperature
	\begin{equation}
		T = \bar{T} q^{-2} = \Oo(q^{-2}),
		\label{very low T scaling relation}
	\end{equation}
	(equivalently $\bar{\beta} = \beta q^{-2}$) which is the regime one must consider to carefully take the zero temperature limit. The chemical potential scaling is defined as $\mu \equiv \bar{\mu} q^2$. In this regime, much like the standard $q$-body SYK model, the ``gaseous'' and ``liquid'' densities diverge \cite{Louw2023Feb}. The terminologies are used in accordance with the van der Waals gas, as we have already seen that both belong to the same universality class. This is highlighted in Fig. \ref{liquid and gas phase}. In fact, the liquid phase, the blue line in Fig. \ref{eosplots} (also see Fig. \ref{liquid and gas phase}), is at $Q_\ell = \Oo(q^0)$ while the gaseous phase has $\Qq_g = \Oo(q^{-1})$. Now recall that the effective SYK interactions have a Gaussian suppression in charge density. As such, the large charge density of the liquid eventually fully suppresses the SYK interactions, leaving a free non-interacting liquid
	\begin{equation}
		\tilde{\Omega}_\ell = \frac{\bar{T}}{2}\ln[1-4\Qq_\ell^2],\quad \bar{\mu}_\ell = 2 \bar{T} \tanh^{-1}(2\Qq_\ell), \label{Freeee}
	\end{equation}
	where recall that we have defined $\bar{\mu} \equiv \mu q^{-2}$.
	
	With this in mind, let us consider the mathematical details around the phase transition in this rescaled regime. The full EOS \eqref{EOS} may be written as
	\begin{equation}
		\bar{\beta}\bar{\mu} =  \ln\left[\frac{1+2\Qq}{1-2\Qq}\right] - 2 \Qq q\bar{\beta}\alpha.
	\end{equation}
	For the case where $\gamma=0$, we know that, at such temperatures and chemical potentials, the phase transition is from $\Qq$ of order $\Oo(1/q) \to \Oo(q^0)$ \cite{Louw2023Feb}. Let us consider the EOS and the grand potential here for such charge densities. As for the $\gamma=0$ case, for charge densities $\Qq = \Oo(q^0)$, the Lyapunov exponent is exponentially suppressed, $v \sim e^{-q \Qq^2} \to 0$, hence the system is free. Here we set $J_q=1$, since we will not focus on the case where $J_q=0$. 
	
	For $\Qq_g = \bar{\Qq}_g q^{-1}$, we find the gaseous phase still remains in the maximally chaotic phase. 
	This leaves us with the gaseous part of the EOS 
	\begin{equation}
		\bar{\mu}_g \sim 4  \bar{\Qq}_g  \sqrt{2 \gamma^2+  1}\label{muFullLowwT}
	\end{equation}
	and the grand potential \eqref{OmegaFull} 
	\begin{equation}
		\tilde{\Omega}_g = -2 \sqrt{1+2 \gamma^2} - 4 \gamma^2 \coth^{-1}  \sqrt{2 \gamma^2+1}.
		\label{OmegaFullLowwwT}
	\end{equation}
	The phase transition will be where \eqref{OmegaFullLowwwT} and \eqref{muFullLowwT} are equal to \eqref{Freeee}. Equating the two expressions, we find that
	$$\Qq_\ell = \frac{\sqrt{1-e^{2\bar{\beta}\tilde{\Omega}_g}}}{2} \xrightarrow[\bar{T} \to 0]{} 1/2$$
	and
	$$ \bar{\Qq}_g   = \bar{T} \frac{\tanh^{-1}(2\Qq_\ell)}{2\sqrt{2 \gamma^2+  1}} \xrightarrow[\bar{T} \to 0]{}  -\frac{\tilde{\Omega}_g}{2\sqrt{1+2\gamma^2}}$$
	
	So for the liquid phase, we have a relatively high density $\Qq_\ell = \Oo(q^{0})$. Recall that such a density leads to exponential suppression in the effective couplings \eqref{effective coupling0} $\Jj_{\kappa q} \lesssim e^{-\kappa q \Qq_{\ell}^2}J_{\kappa q}$. As such, despite the linear $q$ growth stemming from the smaller temperature in $K,U$, defined in Eq. \eqref{U and K defs} or equivalently in Eq. \eqref{U and K defs2}, we find using the scaling relations \eqref{scaling2} and \eqref{very low T scaling relation} that, in this liquid regime, both $K$ and $U$ are smaller than $q\bar{\beta} e^{-q \Qq^2}$. The exponential suppression dominates over any polynomial growth in $q$, implying that the closure relation \eqref{closure2}  
	\begin{equation}
		\pi v = \sqrt{\frac{U^2}{q^2} + \left(\frac{K^2}{\pi v}\right)^2 } q^2 \cos(\pi v/2) + \frac{q^2 K^2}{\pi v}, 
		\label{very low temperature closure relation}
	\end{equation}
	reduces to $\pi v\sim 0$. 
	
	One might consider the case that such an additional rescaling of temperature as in Eq. \eqref{very low T scaling relation} might invalidate the leading order in $q$ result given by $g(t)$. By this, we mean that writing the Green's function $\Gg(t) \propto e^{g(t)/q + g_2(t)/q^2 + \Oo(1/q^3)}$, the question is whether $g(t)$ and $g_2(t)$ can both contribute to the leading order results in the equation of state or the grand potential. At least for the single SYK dot, with $q$-body interactions, this was shown not to be the case. This was shown in appendix A of \cite{Louw2023Feb}. In particular, for any polynomial temperature $T = \Oo(q^\alpha)$, where $\alpha\le 0$, only the leading order correction $g(t)$ contributes to the thermodynamics. Naturally, just because this is true in a single large-$q$ SYK model does not guarantee that the same will hold for our coupled system. The study of this correction term would be one on its own, and whether it has any interesting effects on the thermodynamics is left for a future work.  
	
\begin{figure}[ht]
		\centering
		\includegraphics[width=\columnwidth]{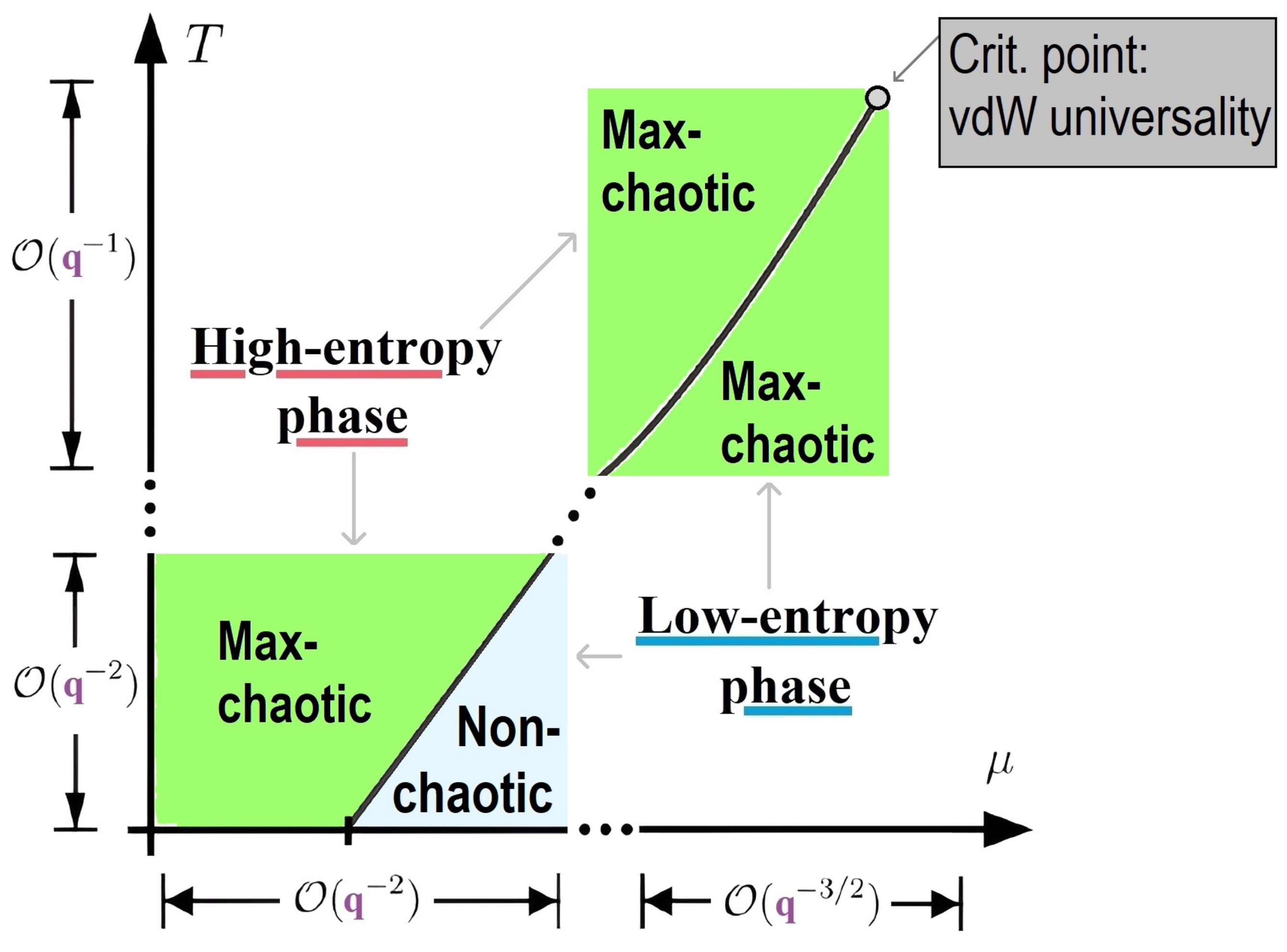}
		\caption{Phase diagram for our coupled SYK system for any finite value of $\gamma$ at low and ``very'' low temperatures, as discussed in section \ref{very low temperatures}. See the text below, Eq. \eqref{very low temperature closure relation} for the assumption involved in the ``very'' low-temperature regime that is left as future work.}
		\label{phase diagram for all T}
	\end{figure}

	We have plotted the phase diagram for low and ``very'' low temperatures in Fig. \ref{phase diagram for all T} for any finite value of $\gamma$. As such, we have also confirmed that the free liquid state is indeed regular (non-chaotic), with a Lyapunov exponent of $\lambda_L = 0$. As such, we are left with a maximally chaotic to regular phase transition, which is similar to that of the standard Hawking-Page transition from a large black hole to non-interacting radiation \cite{Hawking1982Jan} but as stated in the introduction, instead of having a particular value of temperature like in Hawking-Page, we have an entire region of temperatures in the ``very" low-temperature regime in our coupled SYK case where we observe maximally chaotic to non-chaotic first order phase transition. We discuss this in the next subsection.
	
	\subsection{Comparing with the Hawking-Page transition}
	\label{hawking page section}

	\begin{figure}
		\centering
		\includegraphics[width=0.75\columnwidth]{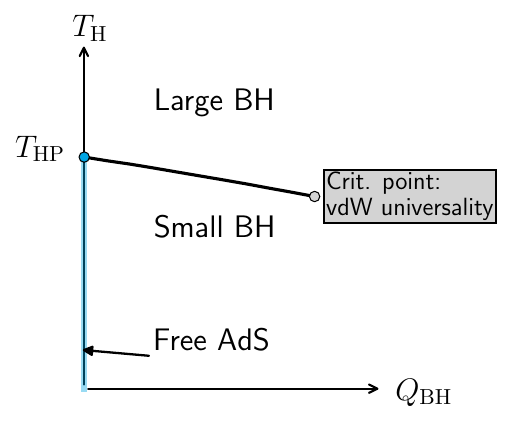}
		\caption{Phase diagram for charged AdS black hole where $T_H$ is the Hawking temperature, $Q_{\text{BH}}$ is the total charge of the black hole and $T_{\text{HP}}$ is the Hawking-Page temperature.}
		\label{phase diagram of black hole}
	\end{figure}
	
	We return to the remark made in the introduction (section \ref{introduction}) where we compare our observed phase transition (see the phase diagram in Fig. \ref{phase diagram for all T}) with the Hawking-Page transition. Following the lead of \cite{Chamblin1999Aug,Chamblin1999Oct}, we present the phase diagram of a charged AdS black hole in Fig. \ref{phase diagram of black hole}. As clear from the figure, the charge-less ($Q_{\text{BH}} = 0$) case has a Hawking-Page transition where, for temperature $T< T_{\text{HP}}$, we have a free (non-chaotic) radiation while a black hole (maximally chaotic) is preferred for temperatures $T> T_{\text{HP}}$. This is a first order phase transition. For all nonzero values of $Q_{\text{BH}} \neq 0$, we have a maximally chaotic large black hole to maximally chaotic small black hole. Contrast this with Fig. \ref{phase diagram for all T} where for low temperatures ($T = \Oo(q^{-1})$), we observe the same behavior as in Fig. \ref{phase diagram of black hole} but for ``very'' low temperatures ($T = \Oo(q^{-2})$), we obtain an entire regime of maximally chaotic to non-chaotic first order phase transition in contrast to Fig. \ref{phase diagram of black hole} where this happens only along the vertical-axis where $Q_{\text{BH}} = 0$. This is what we meant in the introduction and at the end of the previous subsection that for our coupled SYK system, just like single SYK model, due to the presence of this entire range of ``very'' low temperatures, the mapping on the black hole side to a standard Hawking-Page transition is not reproduced. 
	
	Before we end this section, we would like to comment on Fig. \ref{phase diagram of black hole} which we reproduce from \cite{Chamblin1999Aug,Chamblin1999Oct}. The analysis was reproduced and generalized to the charged case in \cite{Kubiznak2012Jul} where the authors called the temperature of transition as the ``Zorro's'' temperature when $Q_{\text{BH}} = 0$ on the coexistence line. As highlighted in appendix \ref{zorro}, we show that this temperature is indeed the same as Hawking-Page temperature and the horizontal-axis of Fig. 13 in \cite{Kubiznak2012Jul} (as reproduced with more details here in Fig. \ref{phase diagram of black hole} taken from \cite{Chamblin1999Aug}, \cite{Chamblin1999Oct}) is the standard Hawking-Page first order phase transition from a maximally chaotic black hole to non-chaotic free radiation (recall this is true only when $Q_{\text{BH}} = 0$) as one cools the temperature down to the Hawking-Page temperature $T_{\text{HP}}$ which is referred to as the ``Zorro's'' temperature in \cite{Kubiznak2012Jul}. We remind the readers again that the ``very'' low-temperature regime in the phase diagram is based on the assumption that only $g(t)$ contributes at leading order at the additional rescaling of temperature $T = \overline{T}q^{-2}$ where we write the Green's function as $\Gg(t) \propto e^{g(t)/q + g_2(t)/q^2 + \Oo(1/q^3)}$. See the text below Eq. \eqref{very low temperature closure relation} where this point is discussed in detail. That is why we have provided these preliminary calculations in the conclusion \& outlook section of our manuscript (i.e., sections \ref{very low temperatures} and \ref{hawking page section}) where we provide a pathway to make these connections robust as the future work.
	
	\subsection{Outlook}
	
	The coupled SYK system's adherence to the established Landau-Ginzburg universality class and its demonstration of maximal chaos are not merely mathematical curiosities. They provide a critical lens through which the fundamental nature of quantum chaos and gravity can be examined. This universality across different models and configurations raises the question of how one can obtain deviating results. Given that we flow to the identical grand potential scaling relations near a critical point for the combination of two SYK models, the natural question to ask is to what extent this result can be generalized. Does it hold for any system of coupled large-$q$ type SYK models with the Hamiltonian given by
	\begin{equation}
		\Hh = \sum_{\kappa > 0} J_{\kappa}\Hh_{\kappa q}
		\label{coupled syk}
	\end{equation}
	or does it extend beyond this, or is it just a coincidence that we found for our coupled case? Given that the kinetic SYK model is integrable, one would imagine this to change upon adding a quadratic term to the above Hamiltonian. This term would dominate in the low-temperature regime, which would at least mean that the model is unlikely to have a maximally chaotic to maximally chaotic second order phase transition, as observed in this work. But without any kinetic term, the future work can be to find if at large-$q$ and low-temperature limits, all coupled SYK systems as in Eq. \eqref{coupled syk} belong to the same universality class as Landau-Ginzburg and if they are also maximally chaotic. This naturally seems plausible from the renormalization group flow arguments were at lower temperatures, the interaction with fewer fermions will dominate the Hamiltonian, but a full analysis for the aforementioned Hamiltonian is yet to be done (see \cite{anninos2023renormalisation} for the Majorana SYK case whose generalization to the complex case will serve as an interesting inquiry). 
	
	Looking ahead, our study naturally leads to the exploration of holographic mappings of these coupled systems to $1+1$-dimensional gravity, analogous to what has been achieved for the single SYK model \cite{Louw2023Oct}. As we have found that the scaling relations near the critical point are the same as the single SYK model, essentially what we have to do is to match our grand potential of the coupled SYK system to that of the single SYK model. Then we use the dictionary developed in \cite{Louw2023Oct} to the deformed JT gravity to find the necessary modifications on the gravity side. We expect to find a phase transition on the gravity side between a small and a large black hole (for example, see \cite{Kubiznak2012Jul}) that corresponds to the chaotic-to-chaotic continuous phase transition as obtained for our model. More importantly, the modifications to be done on the SYK side to reproduce an exact Hawking-Page type first order phase transition, as discussed in detail in section \ref{hawking page section}, will be quite an interesting and welcoming result. We leave these for the future work.

	\section*{Acknowledgments}	
	
	JCL and RJ would like to thank Deutsche Forschungsgemeinschaft (DFG, German Research Foundation))-217133147/SFB 1073 (Project No. B03) for supporting this work. LMvM gratefully acknowledges the support of the Volkswagen foundation, and by the Deutsche Forschungsgemeinschaft (DFG) under
	Grant No 406116891 within the Research Training
	Group RTG 2522/1.
	
	\bibliography{coupled.bib}

	\appendix
	
	\section{Equivalence of our model with a coupled SYK chain at equilibrium}
	\label{equivalence of syk chain with syk dot}
	In this appendix, we prove the equivalence of our model considered in Eq. \eqref{model} with that of a SYK chain with nearest-neighbor hopping. 
	
	We consider a chain of complex SYK dots as studied in Ref. \cite{Jha2023} but with uniform, site-independent coupling whose Hamiltonian at equilibrium (in order to be able to study the thermodynamics of the system) is given by
	\begin{equation}
		\Hh(t) = \sum_{i}  \left(\Hh_{i}(t) + \Hh_{i \to i+1}(t) + \Hh_{i \to i+1}^\dag(t)\right) 
		\label{HamChain}
	\end{equation}
	where the on-site large-$q$ complex SYK Hamiltonian is given by
	\begin{equation}
		\begin{aligned}
			\Hh_{i}(t)
			&= J \hspace{-1mm} \sum\limits_{\substack{ \{\bm{\mu}\}_1^{q/2} \\ \{\bm{\nu}\}_1^{q/2} }} \hspace{-2mm}X(i)^{\bm{\mu}}_{\bm{\nu}} c^{\dag}_{i;\mu_1} \cdots c^{\dag}_{i; \mu_{q/2}} c_{i;\nu_{q/2}}^{\vphantom{\dag}} \cdots c_{i;\nu_1}^{\vphantom{\dag}}
		\end{aligned}
		\label{hi}
	\end{equation}
	summing over $\{\bm{\nu}\}_{1}^{q/2} \equiv 1\le \nu_1<\cdots< \nu_{q/2}\le\Nn$. The transport of $r/2$ fermions from site $i$ to $i+1$ is given by
	\begin{equation}
		\begin{aligned}
			\Hh_{i \rightarrow i+1}(t) 
			&= D \hspace{-1mm} \sum\limits_{\substack{ \{\bm{\mu}\}_1^{r/2} \\ \{\bm{\nu}\}_1^{r/2} }} \hspace{-2mm}Y(i)^{\bm{\mu}}_{\bm{\nu}} c^{\dag}_{i+1;\mu_1} \cdots c^{\dag}_{i+1; \mu_{\frac{r}{2}}} c_{i;\nu_{\frac{r}{2}}}^{\vphantom{\dag}} \cdots c_{i;\nu_1}^{\vphantom{\dag}}.
		\end{aligned}
		\label{hi to i+1}
	\end{equation}
	where $r=q/2$ for our case. Here $X$ and $Y$ are random matrices drawn from Gaussian ensemble with zero mean and variance as in Eq. \eqref{variance of X random matrix} for $\kappa = 1$ and $\kappa = 1/2$, respectively. At equilibrium, there is no heat or charge flow, meaning that each dot has equal temperature and charge. This leads to the Green’s functions also being independent of site label $i$. Accordingly, the self-energy of this system of coupled complex SYK dots with $r=q/2$-body hopping becomes (see Ref. \cite{Jha2023} for more details)
	\begin{equation}
		\begin{aligned}
			q\Sigma_{i}^>(t_1,t_2) =& \hspace{1mm}2 J^2 [-4\Gg_{i}^>(t_1,t_2)  \Gg^<(t_2,t_1)]^{q/2-1} \Gg^>(t_1,t_2) \\
			&+  4|D|^2 [-4\Gg^>(t_1,t_2) \Gg^<(t_2,t_1)]^{r/2-1} \Gg^>(t_1,t_2) \\
			q\Sigma_{i}^<(t_2,t_1) =&\hspace{1mm} 2 J^2  [-4\Gg^>(t_1,t_2)   \Gg^<(t_2,t_1) ]^{q/2-1} \Gg^<(t_2,t_1) \\
			&+  4|D|^2 [-4\Gg^>(t_1,t_2) \Gg^<(t_2,t_1)]^{r/2-1} \Gg^<(t_2,t_1) \\
		\end{aligned}
		\label{se gtrless}
	\end{equation}
	
	We now use the functional form of Green's function 
	\begin{equation}
		\Gg^{\gtrless}(t_1,t_2)  = \mp \left( \frac{1}{2} \mp \Qq \right) e^{g^\gtrless(t_1,t_2)/q} 
		\label{large q gf}
	\end{equation} 
	where $g^{\gtrless}$ is defined in Eq. \eqref{def of g plus minus}. Then plugging Eq. \eqref{large q gf} in Eq. \eqref{se gtrless}, we get the same self-energy for our model in Eq. \eqref{model} as highlighted in eqs. \eqref{ll} and \eqref{self energy in terms of L} below with the identifications $J^2 \to J_q^2$ and $2|D|^2 \to J_{q/2}^2$.
	
	Therefore, from a thermodynamic point of view, our complex SYK dot in Eq. \eqref{model} is the same as an equilibrium chain of complex SYK dots with nearest-neighbor hopping with uniform, site-independent coupling. This is the justification for using the word ``coupled'' in the title of this work. 
	
	\section{Derivation of differential equation for Green's function}
	\label{derivation of differential equations for gs}
	
	For our model in Eq. \eqref{model}, the self-energy $\Sigma$ here is just the sum of the individual self-energies, $\Sigma = \Sigma_{q} + \Sigma_{q/2}$ where $\Sigma_{\kappa q}$ ($\kappa = \{1/2, 1 \}$) is defined below \cite{Maldacena2016Nov}. This enters into the Dyson equation $\Gg^{-1} = \Gg_0^{-1} - \Sigma$, where $\Gg_0$ is the free Green's functions.
	
	We start with the definition of effective coupling coefficients in Eq. \eqref{effective coupling0}. This allows us to define another quantity 
	\begin{equation}
		\Ll^{>}(t_1,t_2) \equiv \hspace{-2mm} \sum_{\kappa \in \{ \frac{1}{2}, 1 \}} \hspace{-2mm} 2\Jj_{\kappa}^2 e^{\kappa g_+(t_1,t_2)}, \Ll^{<}(t_1,t_2) = \Ll^{>}(t_1,t_2)^*
		\label{ll}
	\end{equation}
	where we reproduce the definition of the effective couplings for the Hamiltonian in Eq. \eqref{model} as
	\begin{equation}
		\Jj_{\kappa q} \equiv (1-4\Qq^2)^{\kappa q/4 - 1/2}J_{\kappa q}
		\label{effective coupling}
	\end{equation}
	
	This contributes to the self-energy \cite{Louw2022} of both terms in the Hamiltonian, as
	\begin{equation}
		\Sigma_{\kappa q}^{\gtrless}(t_1,t_2) = \frac{1}{q}\Ll^{\gtrless}(t_1,t_2) \Gg^{\gtrless}(t_1,t_2)
		\label{self energy in terms of L}
	\end{equation}
	
	We are interested in obtaining the differential equations for $g^{\pm}$ whose solutions will solve our model. We start with the differential equations for $g^{\gtrless}$ (plugging Eq. (20) into Eq. (19) in \cite{Louw2022}) where the Kadanoff-Baym equations reduce to the following:
	\begin{equation}
		\partial_{t_1} g^{\gtrless}(t_1,t_2) = \int_{t_1}^{t_2} dt_3  \Ll^{\gtrless}(t_1,t_3) + \underbrace{2\Gg^{\lessgtr}(0,0)}_{2\Qq \pm 1} \i \alpha(t_1).
	\end{equation}
	
	Here $\alpha$ is related to the expectation value of energy per particle given by
	\begin{equation}
		(1-\Qq^2)\alpha(t) = \epsilon_{q/2}(t)/2 + \epsilon_{q}(t) \label{alpha energy}
	\end{equation}
	where in the large-$q$ limit, we have $\Qq = \Oo(q^{-1/2})$ and $\epsilon_{\kappa q}(t)$ is defined as
	\begin{equation}
		\epsilon_{\kappa q}(t) \equiv q^2\frac{J_{\kappa q}\ex{\Hh_{\kappa q}}}{N} 
	\end{equation} 
	where $\kappa \in \{\frac{1}{2}, 1 \}$.
	
	We take the differential equation for $g^< (t_1,t_2)$ and take its complex conjugate to get the differential equation for  time reversed $g^<(t_2, t_1)$ given by (recall $ \Ll^{<}(t_1,t_2) = \Ll^{>}(t_1,t_2)^*$)
	\begin{equation}
		\p_{t_1} g^{<}(t_2,t_1) = \nint[t_1][t_2]{t_3} \Ll^{>}(t_1,t_3) - \left( 2\Qq - 1 \right) \i \alpha(t_1)
	\end{equation}
	
	We add and subtract this to the differential equation for $g^> (t_1, t_2)$ to get 
	\begin{equation}
		\begin{aligned}
			\p_{t_1} g^{+}(t_1,t_2) &= \int_{t_1}^{t_2} dt_3 \Ll^{>}(t_1,t_3) + \i \alpha(t_1) \\
			\p_{t_1} g^{-}(t_1,t_2) &=  2\Qq \i \alpha(t_1)
		\end{aligned}
		\label{first order DE for g}
	\end{equation}

	We recall that we are only interested in the equilibrium situation, implying that all expectation values are constant, hence $\alpha(t)$ is constant as well as the Green's functions are only dependent on time differences $g^{\pm}(t_1,t_2) \equiv g^{\pm}(t_1-t_2)$. Therefore, we solve the above differential equation for $g^-$ to get
	\begin{equation}
		g^{-}(t_1,t_2)= 2\Qq \i \alpha \cdot (t_1-t_2) 
		\label{solution for g-minus}
	\end{equation}
	and obtain a second order differential equation for $g^+$ given by
	\begin{equation}
		\p_{t_1} \p_{t_2}g^{+}(t_1-t_2) = \Ll^{>}(t_1,t_2)
	\end{equation}
	We finally plug the definition of $\Ll^>$ from Eq. \eqref{ll} above to get the differential equation for $g^+$ (where $t_1 - t_2 \equiv t$)
	\begin{equation}
		\ddot{g}^{+}(t) = -2\Jj_q^2 e^{g_+(t)} - 2\Jj_{q/2}^2 e^{g_+(t)/2}
	\end{equation}
	
	Thus we have solved for $g^-$ (Eq. \eqref{solution for g-minus}) and obtained the differential equation for $g^+$ whose solution is presented in the main text in Eq. \eqref{solution for g}, thereby solving the system. 
	
	\section{Energy contributions}
	\label{energy contri}
	
	The energy contributions in Eq. \eqref{alpha energy} can also be derived by the integral expression for $\alpha(t)$ \cite{Jha2023}. Asymptotically in the large-$q$ limit using Eq. \eqref{GreenDef1} and the scaling of $\Qq$ mentioned in Eq. \eqref{scaling1}, we get to leading order in $1/q$ (see eqs. (19) to (20) in \cite{Louw2022})
	\begin{equation}
		\alpha(t_1) = \i \nint[-\infty][t_1]{t_3} \left( q\Sigma^>(t_1-t_3)-q\Sigma^<(t_1-t_3) \right)
	\end{equation}
	where using Eq. \eqref{self energy in terms of L}, we have
	\begin{equation}
		q\Sigma_{\kappa q}^{>}(t_1,t_2) \sim -\Ll^{>}(t_1,t_2)/2 
	\end{equation}
	and $\Ll^{<}(t_1,t_2) = \Ll^{>}(t_1,t_2)^*$, leaving
	\begin{equation}
		\alpha(0) = \Im\nint[-\infty][0]{t} \Ll^{>}(-t)
	\end{equation}
	Note from Eq. \eqref{alpha energy} that $\Qq^2 \sim \Oo(1/q)$ and therefore to leading order in $1/q$, we have $\alpha$ equal to the weighted energy sum (which is also a constant as we are considering equilibrium situation) given by
	\begin{equation}
		\alpha(0) = \sum_{\kappa = \frac{1}{2}, 1} \kappa \epsilon_{\kappa q}(0)
	\end{equation}
	Therefore, we can isolate the individual energy integrals as
	\begin{equation}
		\kappa \epsilon_{\kappa q}(0) = \Im\nint[-\infty][0]{t} 2 \Jj^{2}_{\kappa q} e^{\kappa g^+(-t)} \label{kappaEn}
	\end{equation}

	We first evaluate $\epsilon_{q/2}$ where we substitute Eq. \eqref{solution for g} for $e^{g^+(-t)}$ to get
	\begin{equation}
		\begin{aligned}
			\beta\epsilon_{q/2}(t_1) &=  \Im \nint[-\infty][0]{t} 4\beta^2 T \Jj_{q/2}^2  e^{g^+(-t)/2} \\
			&=  \Im \nint[-\infty][0]{x} \frac{4 \pi v}{1 + B \cos(\pi v/2 +\i x))}
		\end{aligned}
	\end{equation}
	where we have let $x = \pi v T t$ and $B = \sqrt{A^2 + 1}$. Recall that $A$ is given by Eq. \eqref{A2}
	\begin{equation}
		A = \frac{\pi v U}{q K^2}. \label{A3}
	\end{equation}
	The integral is solved to get
	\begin{equation}
		-\frac{8\i \pi v}{A} \tanh^{-1} \left[ \left(\sqrt{A^{-2} +1}-A^{-1}\right) \tan(\pi v/4+\i x/2)\right] 
	\end{equation}
	where $x \rightarrow - \infty$ is a purely real contribution thereby dropping out (since we are interested in the imaginary part of the integral only) and at $x=0$, we are left with
	\begin{equation}
		-\frac{8\i \pi v}{A} \tanh^{-1} \left[ \left(\sqrt{A^{-2} +1}-A^{-1}\right) \tan(\pi v/4)\right] 
	\end{equation}
	Therefore we have the final expression as
	\begin{equation}
		\beta\epsilon_{q/2}(t_1) =  -\frac{8\pi v}{A} \tanh^{-1} \left[ \left(\sqrt{A^{-2} +1}-A^{-1}\right) \tan(\pi v/4)\right] 
		\label{fullEpqo2}
	\end{equation}
	In the limiting case of $\Jj_q\rightarrow 0$, we are left with $A\rightarrow 0$, thereby leading to
	\begin{equation}
		\beta\epsilon_{q/2}(t_1) = -4 \pi v \tan(\pi v/4)
		\label{17}
	\end{equation}
	which is the correct energy density for single large-$q/2$ complex SYK model. Furthermore, considering the (scaled) low-temperature case
	\begin{equation}
		T = \ttil/q = \Oo(q^{-1}),\quad \beta = q \tilde{\beta},
		\label{scaling4}
	\end{equation}
	we find $A  = \Oo(q^{-1})$ implying that the obtained result is still the correct leading order result.

	Next we proceed to calculate $\epsilon_{q}$. We know that to the leading order in $1/q$, we have $\alpha = \epsilon_{q}+\frac{1}{2}\epsilon_{q/2}$. Here we already know $\alpha$ from Eq. \eqref{alpha at low T} which is at low-temperature and $\epsilon_{q/2}$ from Eq. \eqref{fullEpqo2}. Therefore we get
	\begin{equation}
		\begin{aligned}
			E &= \ex{\Hh/N} = \epsilon_{q/2}/q^2 + \epsilon_{q}/q^2 \\
			&= \epsilon_{q/2}/q^2 + (\alpha -\epsilon_{q/2}/2)/q^2
		\end{aligned}
	\end{equation}
	which can be re-arranged as follows:
	\begin{equation}
		q\beta E = \beta\epsilon_{q/2}/(2 q) + \beta\alpha/q
	\end{equation}
	where using the scaling for $\beta$ in Eq. \eqref{scaling4}, we immediately see that $\alpha$ and $q \beta E$ are both at order $\Oo(q^0)$. Then we substitute eqs.  \eqref{alpha at low T} and \eqref{fullEpqo2} to finally get the full expression for the energy in the (scaled) low-temperature limit as
	\begin{equation}
		q\beta E = -\frac{4 K^2}{ U} \coth^{-1} \left[ \sqrt{2 \frac{K^2}{U^2}+1}\right] - 2 U\sqrt{2 \frac{K^2}{U^2} + 1}
		\label{EnergyApp}
	\end{equation}
	
	\section{Chaos theory of the SYK model}\label{app:chaos}
	
	In chaos theory, we wish to analyze the evolution of the system's operator $W(t)$ after a small perturbation $V(0)$ in the initial conditions. The correlation function describing this evolution is the out-of-time-order correlator (OTOC), generally written as
	\begin{equation}
		c(t) = \ex{[W(t), V(0)]^2}_{\beta},
	\end{equation}
	where $\ex{\dots}_{\beta}$ is the thermal expectation value, which can be written as $\ex{\Oo}_{\beta} = \Tr[\rho(\beta) \Oo] = \frac{1}{Z} \Tr[e^{\beta H} \Oo]$. It has been noticed that the correlator grows exponentially with some growth rate $\lambda_L$ \cite{Kitaev2015},
	\begin{equation}
		c(t) \sim e^{\lambda_L t},
	\end{equation}
	and the goal is to find this grow rate. The exponential growth of the OTOC depends on real time, and thus we wish to compute the correlator in real time.

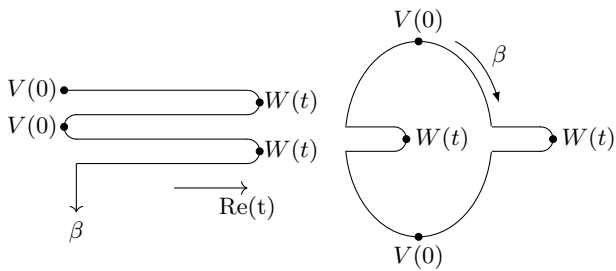
\begin{figure}[h!]
		\centering
		\begin{tikzpicture}[scale=0.65]
			\draw (-0.25,1.5)--(3.5,1.5);
			\draw (0,1)--(3.5,1);
			\draw (0,0.5)--(3.5,0.5);
			\draw (0,0)--(3.5,0);
			\draw[->](0,0)--(0,-1) node[anchor=north]{$\beta$};
			\node at (4.4,0.25){$W(t)$};
			\node at (4.4,1.25){$W(t)$};
			\node at (-0.9,0.75){$V(0)$};
			\node at (-0.9,1.5){$V(0)$};
			\draw[fill](3.75,0.25) circle (2pt);
			\draw[fill](-0.25,0.75) circle (2pt);
			\draw[fill](3.75,1.25) circle (2pt);
			\draw[fill](-0.25,1.5) circle (2pt);
			\begin{scope}
				\clip (3.5,0) rectangle (4,1.5);
				\draw (3.5,0.25) circle (0.25);
				\draw (3.5,1.25) circle (0.25);
			\end{scope}
			\begin{scope}
				\clip (0,0.5) rectangle (-0.5,1);
				\draw (0,0.75) circle (0.25);
			\end{scope}
			%\draw[->](-1.5,0.5)--(-1.5,2) node[anchor=south]{Im(t)};
			\draw[->](2,-0.5)--(3.5,-0.5) node[anchor=north]{Re(t)};
			
			\draw (7,0.5) circle (1.5 and 2);
			\filldraw[white](5.3,0.25) rectangle (5.7,0.75);
			\filldraw[white](8.3,0.25) rectangle (8.7,0.75);
			\draw (5.5,0.75) -- (6.5,0.75);
			\draw (5.5,0.25) -- (6.5,0.25);
			\draw (8.5,0.75) -- (9.5,0.75);
			\draw (8.5,0.25) -- (9.5,0.25);
			
			\begin{scope}
				\clip (6.5,0.25) rectangle (7,0.75);
				\draw (6.5,0.5) circle (0.25);
			\end{scope}
			\begin{scope}
				\clip (9.5,0.25) rectangle (10,0.75);
				\draw (9.5,0.5) circle (0.25);
			\end{scope}
			\draw[-latex] (7.75,2.5) arc
			[
			start angle=70,
			end angle=20,
			x radius=1.5,
			y radius =2.1
			] ;
			\node at (8.65,2.2) {$\beta$};
			\filldraw[black] (6.75,0.5) circle (2pt) node[anchor = west]{$W(t)$};
			\filldraw[black] (9.75,0.5) circle (2pt) node[anchor = west]{$W(t)$};
			\filldraw[black] (7,2.5) circle (2pt) node[anchor = south]{$V(0)$};
			\filldraw[black] (7,-1.5) circle (2pt) node[anchor = north]{$V(0)$};
			
		\end{tikzpicture}
		\caption{On the left is the Keldysh-Schwinger contour for the out-of-time-correlator (OTOC) for the chaotic system. The perturbation $V(0)$ and generic operator $W(t)$ are separated by a large real time and a quarter on the thermal circle, as is shown in the picture on the right.}
		\label{fig:KS contour}
	\end{figure}

	A common and convenient \emph{choice} of writing the OTOC in SYK models is \cite{MSS2016}
	\begin{equation}
		\mathcal{F}(t_1,t_2) = \Tr\left[y \;W(t)\; y \; V(0) \;y \;W(t)\; y\; V(0)\right], \quad y\equiv \rho(\beta)^{1/4}
	\end{equation}
	where $\rho(\beta)$ is the thermal density matrix. This choice is to make computation easier, and will not affect the exponent \cite{MSS2016, Stanford2016}. For simplicity, we have illustrated the OTOC with a Keldysh-Schwinger contour in figure (\ref{fig:KS contour}). 
	
	\FloatBarrier
	
	Here, the operators are separated by a quarter on the thermal circle (for details, see \cite{Maldacena2016Nov}), and additionally a real-time separation between the pair of operators. We are interested in regimes where $t \rightarrow \infty$, such that the zero rung diagrams, can be neglected.  The paths in real time are described by the retarded propagator defined as (\ref{retard and wightman}), and the Wightman Greens function (\ref{retard and wightman}) describes a pair of operators separated by $\beta/2$ on the thermal circle. If we evaluate the Greens function (\ref{solution for g}) with $t \rightarrow it+\beta/2$, we see that the function becomes a function of real time. Hence, this operation can be seen as an analytic continuation to real time.

	%Now that we have the propagators, we can compute the retarded kernel $K_R$ and construct the $\Oo{\frac{1}{N}}$ part of $\mathcal{F}$. In his talk \cite{Kitaev2015}, Kitaev points out that the OTOC grows exponentially, and thus we make the ansatz
	
	%$$\mathcal{F}(t_1,t_2= e^{\lambda_L (t_1+t_2)} f(t_{12}).$$

	\begin{figure}[ht]
		\centering
		\includegraphics[width=0.75\columnwidth]{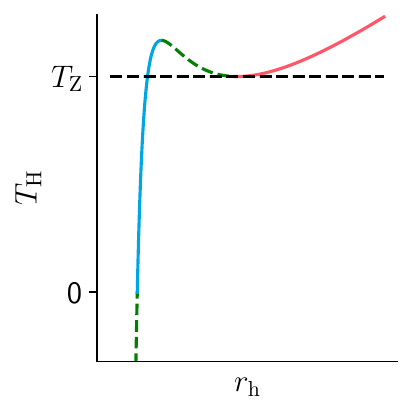}
		\caption{Equation of state for charged AdS black hole where $T_H$ is the Hawking temperature, $Q_{\text{BH}}>0$ is the total charge of the black hole and $T_{\text{Z}}$ is the minimum temperature of the larger black hole. The green dashed lines are the unstable regimes with negative specific heat. The two stable phases are marked with blue and red. Color coding is the same as in Figs. \ref{eosplots}, \ref{Phases} and \ref{liquid and gas phase}.}
		\label{EOS of RN black hole}
	\end{figure}

	\section{``Zorro's'' temperature and the Hawking-Page transition}
	\label{zorro}
	For a Reissner–Nordstr\"om black hole, we find the Hawking temperature
	\begin{equation}
		T_{\text{H}} = \frac{1}{4\pi r_\text{H}} \left(1+8
		\pi r_\text{H} P - \frac{3 Q_{\text{BH}}}{r_\text{H}^2}\right) \label{TTApf}
	\end{equation} 
	where $r_\text{H}$ is the horizon radius, $P = -\Lambda/(8\pi)$ is a pressure term stemming from the negative cosmological constant $\Lambda$, so $P>0$, and $Q_{\text{BH}}$ is the charge of the black hole. 
	
	Here, we partially repeat the analyses done in \cite{Chamblin1999Aug,Chamblin1999Oct,Kubiznak2012Jul}. The goal is to provide a simple picture for how the standard large-to-small charged black hole phase transition reduces to the Hawking-Page transition \cite{Hawking1982Jan} at zero charge. 
	
	While it is true that the phase diagram is defined by both the thermodynamic potential and the equation of state, one may extract much information from just the latter. This is because the unstable phases are highlighted by a negative specific heat $C = T_{\text{H}} \p_{T_{\text{H}}} S$, with entropy $S = \pi r_{\text{H}}^2$. Using \eqref{TTApf} we find 
	\begin{equation}
		C = \frac{2 \pi r_{\text{h}}^2(9 r_{\text{h}}^4 + 8 \pi P(r_{\text{h}}^2-Q_{\text{BH}}^2))}{9 r_{\text{h}}^4 - 8 \pi P(r_{\text{h}}^2-3Q_{\text{BH}}^2)}
	\end{equation}
	
	We have plotted the EOS for various horizon radii in Fig. \ref{EOS of RN black hole}. The phases with negative specific heat are given by the dotted lines. Note that only two stable phases remain, corresponding to the small and large black holes in blue and red, respectively.
	
	It is also important to note that the larger black hole cannot exist below the ``Zorro's'' \footnote{The term ``Zorro's'' temperature was coined in \cite{Kubiznak2012Jul} due to the Z-slash-like 'Zorro' free energy curve in Fig. 5 of \cite{Chamblin1999Oct}.} temperature $T_{Z}(Q_{\text{BH}})$. As such, below this temperature, the larger black hole must undergo a phase transition to the smaller black hole.

	As the black hole's charge decreases, however, the blue line becomes steeper and the range of radii $r_{\text{H}} \in [r_{\text{small}}^{(-)},r_{\text{small}}^{(+)}]$ with
	\begin{align*}
		r_{\text{small}}^{(-)} &= \sqrt{(-1 + \sqrt{1+12[Q_{\text{BH}}8 \pi P/3]^2 })/6}\\
		r_{\text{small}}^{(+)} &= \sqrt{(1 - \sqrt{1-36 [Q_{\text{BH}} 8 \pi P/3]^2})/6}
	\end{align*}
	becomes smaller for the smaller black hole (denoted by the blue line in Fig. \ref{EOS of RN black hole}). It is only at $Q_{\text{BH}}=0$ that this range tends to zero. It is then natural to ask what happens to the larger black hole below $T_{Z}(Q_{\text{BH}}=0)$. For this special case, we find that $T_{Z}(0)$ tends to the Hawking-Page temperature $T_{\text{HP}}$ \cite{Hawking1982Jan} using Eq. \eqref{TTApf}. Further, it is in this case that the non-interacting radiation phase (thermal AdS) becomes the thermodynamically preferred phase, as depicted in Fig. \ref{phase diagram of black hole}.

\end{document}